\documentclass[onecolumn,aps,prd,preprintnumbers,showpacs,superscriptaddress,nofootinbib,amsmath,amssymb,floats,floatfix,showkeys,notitlepage,longbibliography]{revtex4-1}

\usepackage{orcidlink}
\usepackage{lipsum}
\usepackage{graphicx}
\usepackage{subfigure}
\usepackage{braket}
\usepackage[utf8]{inputenc}
\usepackage{sans}
\usepackage{changes}
\usepackage{hyperref}
\hypersetup{colorlinks=true,linkcolor=blue,urlcolor=blue,citecolor=blue}
\usepackage{lmodern}

\usepackage{mathrsfs}
\usepackage{mathtools}
\usepackage[toc,page]{appendix}
\usepackage[normalem]{ulem}
\usepackage{adjustbox}
\usepackage{latexsym}
\usepackage{amsmath}
\usepackage{amssymb}
\usepackage{amsfonts}
\usepackage{dcolumn}
\usepackage{bm}
\usepackage{tikz}
\usepackage{bigints}
\usepackage{array,tabularx,multirow,booktabs}
\usepackage[tracking=true]{microtype}
\SetTracking{}{500}
\SetTracking{encoding={*}, shape=sc}{40}
\UseRawInputEncoding %for inputenc error%
\allowdisplaybreaks

\begin{document} \sloppy

\title{Constraints on Kalb-Ramond Gravity from EHT Observations of Rotating Black Holes in Traceless Conformal Electrodynamics}
%Constraining Kalb-Ramond Gravity with EHT Observations of Rotating Black Holes Coupled to Traceless Conformal Electrodynamics 

\author{Y. Sekhmani\orcidlink{0000-0001-7448-4579} }
\email[Email: ]{sekhmaniyassine@gmail.com}
\affiliation{Center for Theoretical Physics, Khazar University, 41 Mehseti Street, Baku, AZ1096, Azerbaijan.}
%\affiliation{Institute of Nuclear Physics, Ibragimova, 1, 050032 Almaty, Kazakhstan}
\affiliation{Centre for Research Impact \& Outcome, Chitkara University Institute of Engineering and Technology, Chitkara University, Rajpura, 140401, Punjab, India}
%\affiliation{Department of Mathematical and Physical Sciences, College of Arts and Sciences, University of Nizwa, P.O. Box 33, Nizwa 616, Sultanate of Oman}

\author{K. Boshkayev\orcidlink{http://orcid.org/0000-0002-1385-270X}}
\email{kuantay@mail.ru}
\affiliation{Al-Farabi Kazakh National University, Al-Farabi av. 71, 050040 Almaty, Kazakhstan}
%\affiliation{Institute of Nuclear Physics, Ibragimova, 1, 050032 Almaty, Kazakhstan}
%\affiliation{Kazakh-British Technical University, Tole bi str., 59, Almaty, 050000, Kazakhstan}

\author{M. Azreg-A\"{\i}nou\orcidlink{0000-0002-3244-7195}}
\email{azreg@baskent.edu.tr}
\affiliation{Ba\c{s}kent University, Engineering Faculty, Ba\u{g}l{\i}ca Campus, 06780-Ankara, T\"{u}rkiye}

\author{S. K. Maurya\orcidlink{0000-0003-0261-7234}}  \email[Email: ]{sunil@unizwa.edu.om}
\affiliation{Department of Mathematical and Physical Sciences, College of Arts and Sciences, University of Nizwa, P.O. Box 33, Nizwa 616, Sultanate of Oman}

\author{M. Altanji \orcidlink{https://orcid.org/0000-0002-8137-2046}}
\email[Email: ]{maltenji@kku.edu.sa}
\affiliation{Department of Mathematics, College of Sciences, King Khalid University, Abha 61413, Saudi Arabia}

\author{A. Urazalina \orcidlink{https://orcid.org/0000-0002-4633-9558}}
\email{y.a.a.707@mail.ru}
\affiliation{Al-Farabi Kazakh National University, Al-Farabi av. 71, 050040 Almaty, Kazakhstan}
%\affiliation{Institute of Nuclear Physics, Ibragimova, 1, 050032 Almaty, Kazakhstan}
%\affiliation{Kazakh-British Technical University, Tole bi str., 59, Almaty, 050000, Kazakhstan}

\begin{abstract}
We present a phenomenological study of rotating, charged black holes in Einstein gravity coupled to a traceless (conformal) matter sector formed by ModMax nonlinear electrodynamics and a Kalb-Ramond two-form that spontaneously breaks local Lorentz symmetry. Starting from a family of obtained static, Schwarzschild-like solutions with a traceless Kalb-Ramond sector, we construct the stationary, axisymmetric counterpart via the Newman-Janis algorithm. The resulting Newman-Kerr-like metric depends on four intrinsic parameters: the electric charge $Q$, the ModMax nonlinearity $\gamma$, the Lorentz-violation amplitude $\ell$ and the spin $a$. We analyze horizon structure and separatrices in parameter space, derive the null geodesic equations and obtain the photon capture boundary that defines the black hole shadow. Using ray-tracing, we compute shadow silhouettes and a suite of shadow observables (areal radius, characteristic radius $R_s$, distortion $\delta$, oblateness $D$) and show how $\gamma$ and $\ell$ produce qualitatively distinct effects: $\gamma$ acts as a screening factor for the electromagnetic imprint, while $\ell$ introduces angular-dependent metric rescalings that deform shadow shape beyond simple size rescaling. We confront model predictions with EHT angular-radius measurements for M87$^*$ and Sgr A$^*$ and derive conservative bounds on the combinations of $(Q,\gamma,\ell,a)$. Our results identify an effective charge combination $Q_{\rm eff}\simeq e^{-\gamma}Q^{2}/(1-\ell)^{2}$ and demonstrate that modest $Q_{\rm eff}$ remains compatible with current EHT images while large $Q_{\rm eff}$ is progressively disfavored.
\end{abstract}

\maketitle

\section{Introduction}
%The notion of a black hole (BH) \emph{shadow} has become a precise and powerful diagnostic of strong-field gravity: it is the dark silhouette produced when gravitational lensing and photon capture by the event horizon extinguish background emission. Physically, the shadow corresponds to the sky projection of photon-capture cross sections as seen by a distant observer and therefore encodes the spacetime's null-geodesic structure. Early foundational work by Synge and, subsequently, by Bardeen \cite{Synge:1966okc,Bardeen:1973} who computed the apparent shape of the event horizon of a Kerr BH and thereby introduced the modern conception of the shadow for a distant observer-established the conceptual and calculational framework that underpins contemporary observational and theoretical studies. Because the shadow boundary is determined directly by the metric's null geodesics, its morphology is sensitive to the BH's mass, spin, and any deviations from general relativity (GR). 

A black hole is similar to a coin placed beneath a sheet of paper; although it is invisible, its effect on the distribution of light can be observed. The shape and characteristics of the coin remain constant, and whenever you rub the paper, the impression reveals the same features. The spacetime structure of a black hole is inherently fixed. Its influence remains constant, irrespective of the shape, colour, or behaviour of external electromagnetic radiation, provided that a few fundamental principles are adhered to. The existence of the black hole can be revealed through illumination from external sources of electromagnetic radiation.

Recent observations by the Event Horizon Telescope (EHT) Collaboration most notably the published images of the supermassive black holes M87$^{*}$ \cite{EventHorizonTelescope:2019dse,EventHorizonTelescope:2019uob,EventHorizonTelescope:2019jan,EventHorizonTelescope:2019ths,EventHorizonTelescope:2019pgp,EventHorizonTelescope:2019ggy} and Sgr A$^{*}$ \cite{EventHorizonTelescope:2022wkp,EventHorizonTelescope:2022apq,EventHorizonTelescope:2022wok} offer direct visual probes of the strong-field regime of gravity. These images depict an effectively stationary spacetime geometry illuminated by a rapidly varying emission region, thereby demanding a more refined interpretation of both the observable signatures and the information that remains concealed. The BH shadow is a prominent feature of these observations: its angular size and silhouette encode a characteristic imprint of the underlying compact object. Detailed characterization of shadow morphology is therefore essential not only for the identification of black holes, but also for stringent tests of general relativity and its alternatives. Building on the Synge-Bardeen procedure, the literature has systematically explored shadows across a wide range of spacetimes and extensions of GR: quantum-corrected and non-Kerr models \cite{Luminet:1979nyg,Johannsen:2013vgc,Amir:2016cen,Abdujabbarov:2016hnw,Kumar:2017tdw,Shaikh:2019fpu,Gralla:2019xty,Wei:2020ght,Huang:2023yqd,Meng:2023htc,Liu:2020ola,Yang:2022btw,Zhang:2023okw,Afrin:2021wlj}; rotating solutions in modified theories such as Chern-Simons gravity, braneworld and Kaluza-Klein constructions \cite{Amarilla:2010zq,Amarilla:2011fx,Amarilla:2013sj}; Horava-Lifshitz gravity, traversable wormholes and other non-Kerr geometries \cite{Atamurotov:2013dpa,Nedkova:2013msa,Atamurotov:2013sca}; higher-dimensional backgrounds \cite{Papnoi:2014aaa,Singh:2017vfr,Ahmed:2020jic,Singh:2023ops}; and black holes (BHs) coupled to nonlinear electrodynamics \cite{Atamurotov:2015xfa,Kumar:2019pjp,Kumar:2020ltt}. Investigations that incorporate exotic matter (e.g. quintessence) \cite{Abdujabbarov:2015pqp}, regular (nonsingular) metrics \cite{Abdujabbarov:2016hnw}, higher-curvature corrections \cite{Dastan:2016vhb,Abdujabbarov:2016hnw,Afrin:2024khy}, noncommutative geometries \cite{Sharif:2016znp}, quadrupolar deformations \cite{Wang:2017hjl} and asymptotically-safe gravity scenarios \cite{Kumar:2019ohr} further demonstrate how shadow observables furnish diverse and stringent tests of gravitational dynamics in the strong-field regime.
Moreover, shadow studies bear on a range of foundational problems in astrophysics and fundamental physics, including the physics of extreme gravitational environments, the properties and distribution of dark matter, the dynamics and radiative processes of accretion flows, the physical mechanisms behind cosmic acceleration, and the possible phenomenology of extra spatial dimensions. For further examples and extended discussion see \cite{Grenzebach:2014fha,Abdujabbarov:2015xqa,Guo:2018kis,Zhu:2019ura,Long:2019nox,Chowdhuri:2020ipb,Lee:2021sws,Konoplya:2021slg,Zhang:2021hit,Nampalliwar:2021tyz,Junior:2021svb,Qin:2022kaf,Wang:2022kvg,Zeng:2021mok,Zhang:2022klr,Wang:2022mjo,Ghosh:2022mka,Vagnozzi:2022moj,Zubair:2023cor,Galishnikova:2023ltq,Broderick:2023jfl,Zhang:2024jrw,Jiang:2023img,Nguyen:2023clb,Huang:2024wpj,Wang:2023jop,Chen:2023wzv,Raza:2024zkp,Wei:2024cti,Liu:2024lbi,Kuang:2024ugn,Belhaj:2022kek,Belhaj:2023nhq,Fathi:2025ikx,Sekhmani:2025bsi,Yang:2025byw}.

Considering quantum-motivated modifications of gravity is imperative: in particular, studies of Lorentz-symmetry breaking (LSB) furnish a vital window onto possible departures from canonical gravitational dynamics and thus play a central role in attempts to reconcile gravity with fundamental physics. By isolating the low-energy imprints of LSB on the spacetime geometry and by confronting these imprints with observational diagnostics such as black hole shadows, one obtains a sensitive probe of Lorentz-violating effects relative to the predictions of general relativity. Such an approach permits an assessment of whether deviations from Lorentz invariance leave observable signatures in photon propagation, horizon-scale lensing, and the silhouette of compact objects. \cite{Kostelecky:1991ak,Colladay:1998fq,Casana:2017jkc,Ovgun:2018xys,Gullu:2020qzu,Pan:2020zbl,Liu:2022xse,Maluf:2020kgf,Xu:2022frb,Ding:2021iwv,Poulis:2021nqh,Mai:2023ggs,Xu:2023xqh,Zhang:2023wwk,Lin:2023foj,Chen:2023cjd,Chen:2020qyp,Wang:2021gtd,Liu:2024oeq,Mai:2024lgk,Liang:2022gdk,Tian:2021mka,Tian:2022gfa,Hosseinifar:2024wwe,Finke:2024ada,Liu:2024axg}. A paradigmatic realization of Lorentz violation is provided by Kalb-Ramond (KR) gravity, in which an antisymmetric two-form field $B_{ab}$ (the KR field) is nonminimally coupled to the metric. The KR field arises naturally in the bosonic sector of string theory and therefore offers a well-motivated effective-field-theory setting to study LSB-induced modifications of spacetime structure. \cite{Kalb:1974yc,Kao:1996ea} Considerable effort has been devoted to constructing exact solutions within this framework \cite{Chakraborty:2014fva,Maluf:2021ywn,Duan:2023gng,Yang:2023wtu}. Notably, Yang \emph{et al.} \cite{Yang:2023wtu} presented the first correct derivation of a Schwarzschild-like solution in KR gravity and, by relaxing the vacuum conditions, the corresponding Schwarzschild-(A)dS analogue. Building on these results, in a joint paper we identified more general spherically symmetric, neutral solutions in the same theory \cite{Liu:2024oas}, and the ensuing phenomenology of these black holes has since attracted sustained attention in the literature. \cite{Guo:2023nkd,Filho:2023ycx,AraujoFilho:2024rcr,Jha:2024xtr,Junior:2024ety,Junior:2024vdk,AraujoFilho:2024rcr,Du:2024uhd,Filho:2024tgy}

In recent years, there has been renewed interest in a class of nonlinear electrodynamics known as \emph{ModMax} (modified Maxwell) theory \cite{Bandos:2020hgy}. Nonlinear extensions of Maxwell electrodynamics are attractive because they can ameliorate or regulate classical field singularities and open a controlled window onto strong-field electromagnetic phenomena. The theoretical structure and black-hole solutions of ModMax electrodynamics have been examined in considerable detail (see, e.g., \cite{Sorokin:2021tge,Kosyakov:2020wxv,Bandos:2021rqy,Kruglov:2021bhs,Flores-Alfonso:2020euz,BallonBordo:2020jtw,Kubiznak:2022vft,Barrientos:2022bzm,Siahaan:2023gpc,Siahaan:2024ajn,EslamPanah:2024fls,EslamPanah:2024tex} and references therein). A salient feature of ModMax is that the nonlinear coupling introduces an effective screening of the electric and magnetic charges via an exponential factor in the field strength, with important consequences for the near-horizon geometry and asymptotic charges. Remarkably, the exact static, spherically symmetric charged solution in Einstein-ModMax gravity closely parallels the Reissner-Nordstr\"om geometry, differing only through ModMax-induced deformations of the metric functions and the effective charge profile \cite{Flores-Alfonso:2020euz}. Extensions to dyonic configurations and studies of associated observables-shadow morphology, gravitational lensing, and quasinormal-mode spectra have been investigated (see, in particular, \cite{EslamPanah:2024fls}), highlighting the potential for these signatures to become accessible as observational capabilities continue to advance.

In this work we investigate the influence of traceless field contributions and spontaneous LSB on the shadow morphology of rotating black holes, and we derive observational constraints on the associated coupling parameters by numerically estimating the angular radius of the supermassive black holes Sgr~A$^{*}$ and M87$^{*}$. Concretely, we begin from the family of previously obtained Schwarzschild-like solutions endowed with a traceless Kalb-Ramond sector and implement a rotation parameter $a$ by means of the Newman-Janis algorithm (NJA)~\cite{Azreg-Ainou:2014pra,Azreg-Ainou:2014aqa,Azreg-Ainou:regular} to construct the corresponding stationary, axially symmetric geometry. On this rotating background we compute the null geodesic structure and determine the photon capture cross section that defines the shadow boundary as seen by a distant observer. We then produce contour plots of the shadow in an appropriate observational (physical) image plane for a representative set of model benchmarks, varying the charge, the ModMax parameter, the traceless-field amplitude, and the Lorentz-violation couplings. In parallel, we analyze the associated energy-emission (spectral) rate arising from near-horizon processes to assess complementary observational signatures. Finally, by comparing the numerically obtained angular radii and shadow morphologies with the EHT measurements for M87$^{*}$ and Sgr~A$^{*}$, we place quantitative bounds on the model parameters and discuss the degree to which traceless-field effects and spontaneous Lorentz violation are compatible with current horizon-scale imaging. The paper is organized as follows. In Sec.~\ref{sec:background} we summarize the static solutions, in Sec.~\ref{sec:Solution} we derive the static solution and in Sec.~\ref{sec:MNJ} we implement the NJA for obtaining the rotating counterpart. In Sec.~\ref{sec:geodesics} we derive the null-geodesic equations and shadow boundary conditions and in Sec.~\ref{sec:Shadow} we evaluate the shadow observables. Sec.~\ref{sec:results} presents contour plots, emission-rate calculations and parameter scans; and in Sec.~\ref{sec:constraints} we confront our results with EHT data and discuss observational constraints and prospects. We summarize in Sec.~\ref{sec:discussion}. Throughout this paper, we adopt the geometric unit $G=c=1$.

\section{Lorentz-violating gravity with a background KR field}\label{sec:background}

To properly proceed with finding a black hole solution in a self-interacting Kalb-Ramond field, it might be best to first represent the main features of the Kalb-Ramond field. Generally, the Kalb-Ramond field is an antisymmetric tensor with rank two, typically featured by $B_{\mu\nu}=-B_{\mu\nu}$. Moreover, the associated gauge invariant pertinent to the field intensity is a tensor three-form $H_{\mu\nu\rho}\equiv \partial_{[\mu}B_{\nu\rho]}$, which is invariant under the transformation $B_{\mu\nu} \to B_{\mu\nu}+\partial_{[\mu}\Gamma_{\nu]}$. In addition, by introducing a time vector $v^\alpha$, the Kalb-Ramond field can be broken down as follows: $B_{\mu\nu}=\tilde E_{[\mu}v_{\nu]}+\epsilon_{\mu\nu\alpha\beta}v^\alpha \tilde B^\beta$ with $\tilde E_\mu v^\mu=\tilde B_\mu v^\mu=0 $ \cite{Altschul2010, Lessa2020}. Along this decomposition, the spatial pseudo-vectors $\tilde E_\mu$ and $\tilde B_\mu$ act as the pseudo-electric and pseudo-magnetic similarly to Maxwell's electromagnetic fields, respectively; in other words, they reflect the dynamic degrees of freedom of $B_{\mu\nu}$.

The model is described by a minimal coupling of Einstein gravity governed by the Ricci scalar $R$ to a self-interacting Kalb-Ramond two-form $B_{\mu\nu}$ with its quadratic term $B^{\mu\nu}B_{\mu\nu}$; the dynamics are encoded in the following action: \cite{Altschul2010, Lessa2020}
\begin{eqnarray}
\mathcal{S}&=&\frac{1}{2}\!\int{\!d^4x\sqrt{-g}}\bigg[R \!-\! \frac{1}{6}H^{\mu\nu\rho}H_{\mu\nu\rho}\!-\!V(B^{\mu\nu} B_{\mu\nu}\!\pm \!b^2) +\xi_2 B^{\rho\mu}B^{\nu}{}_\mu R_{\rho\nu}+\xi_3 B^{\mu\nu}B_{\mu\nu}R  \bigg] +\int{d^4x\sqrt{-g}\mathcal{L}_{M}},
\label{Main_Action}
\end{eqnarray}
Here, $(\xi_{2}, \xi_{3})$ are dimensionless parameters pertinent to the non-minimal coupling of the quadratic quantity $B_{\mu\nu}B^{\mu\nu}$ to the curvature of spacetime. Thus, upon closer examination of the associated potential $V(B^{\mu\nu}B_{\mu\nu}\pm b^2)$, one can observe that it is sensitive only to the Lorentz-invariant combination $B^{\mu\nu}B_{\mu\nu}$ and is therefore invariant under local (observer) Lorentz transformations. Hence, the potential achieves its minimum only whenever the Lorentz invariant condition $B^{\mu\nu} B_{\mu\nu}=\mp b^2$ is satisfied, guaranteeing the definite positivity of $b^2$. As a result, the Kalb-Ramond field implies a non-zero vacuum expectation value (VEV) $(\langle B_{\mu\nu} \rangle = b_{\mu\nu})$. Concretely, non-minimal coupling implies a vacuum, underlining a spontaneously broken local Lorentz symmetry in the matter sector. Here, the term $\xi_{3}B^{\mu\nu}B_{\mu\nu}R=\mp\xi_{3}b^{2}R$ results in a Ricci scalar and is effectively absorbed into the Einstein-Hilbert term via a redefinition of the gravitational coupling.

A stringent investigation of a black hole solution with a self-interacting Kalb-Ramond field is essentially governed by the matter sector, which in this context comprises a charged source described by traceless conformal electrodynamics. Consequently, the relevant exact Lagrangian for matter can be described as follows: 
\begin{eqnarray}
\mathcal{L}_{\rm M} = \mathcal{L}_{\rm ModMax} + \mathcal{L}_{\rm int},
\end{eqnarray}
where $\mathcal{L}_{\rm ModMax}$ denotes the Lagrangian density of ModMax nonlinear electrodynamics, and $\mathcal{L}_{\text{int}}$ encodes the interaction between the electromagnetic field and the Kalb-Ramond background. To be more precise, the ModMax Lagrangian can be described as follows \cite{Cirilo-Lombardo:2023poc,Kosyakov:2020wxv}:
\begin{equation}
\mathcal{L}_{\text{ModMax}}(\gamma) = -\mathcal{S}\cosh \gamma + (\mathcal{S}^2 + \mathcal{P}^2)^{1/2}  \sinh \gamma
\end{equation}
with
\begin{equation}
\mathcal{S} := \frac 14 F_{\mu\nu} F^{\mu\nu}, \quad\quad                           \mathcal{P} := \frac 14 F_{\mu\nu}\tilde F^{\mu\nu}
\end{equation}
with $F_{\mu\nu}=\partial_{[\mu}A_{\nu]}$ and $\tilde{F}_{\mu\nu}:=\tfrac{1}{2}\epsilon_{\mu\nu\sigma\rho}F^{\sigma\rho}$. Here $\gamma$ is the ModMax parameter: $\gamma=0$ recovers Maxwell electrodynamics, while $\gamma\neq0$ gives rise to birefringence. In that case one polarisation retains the standard lightlike dispersion relation, whereas the other acquires a modified phase velocity which is subluminal for $\gamma>0$ (and would become superluminal for $\gamma<0$). For physical consistency one therefore imposes $\gamma\geq0$.

A hands-on investigation of the ModMax field is supported by Plebanski's dual variable, stated as follows:
\begin{eqnarray}
P_{\mu\nu}=-\mathcal{L}_{\mathcal{S}} F_{\mu\nu} -\mathcal{L}_{\mathcal{P}} \tilde{F}_{\mu\nu}= \left( \cosh \gamma - \frac{\mathcal{S}}{(\mathcal{S}^2 + \mathcal{P}^2)^{1/2}} \sinh \gamma \right) F_{\mu\nu}
-\frac {\mathcal{P} \sinh \gamma}{(\mathcal{S}^2 + \mathcal{P}^2)^{1/2}} \tilde F_{\mu\nu} \label{Pmodmax}
\end{eqnarray}

The dual of $P_{\mu\nu}$ is then given by
\begin{eqnarray}
\tilde P_{\mu\nu} &=& \left( \cosh \gamma - \frac {\mathcal{S}}{(\mathcal{S}^2 + \mathcal{P}^2)^{1/2}} \sinh \gamma \right) \tilde F_{\mu\nu}
+\frac {\mathcal{P} \sinh \gamma}{(\mathcal{S}^2 + \mathcal{P}^2)^{1/2}} F_{\mu\nu}. 
\end{eqnarray}
Effectively, ModMax theory is assumed to be invariant under conformal transformations of the metric, $g \to \Omega^2 g$, and invariant under duality rotations of $SO(2)$. So, one has
\begin{equation}\label{dualityRot}
\begin{pmatrix}
P'_{\mu\nu}\\
\Tilde{F}'_{\mu\nu}
\end{pmatrix}
=
\begin{pmatrix}
\cos\theta & \sin \theta\\
-\sin\theta & \cos\theta 
\end{pmatrix}
\begin{pmatrix}
P_{\mu\nu}\\
\Tilde{F}_{\mu\nu}
\end{pmatrix}\,,
\end{equation}
where $P$ is usually defined as $P=(1/4)P_{\mu\nu}P^{\mu\nu}$, so that
\begin{equation}
P=\Bigl(\cosh \gamma -\frac{{\cal S}\sinh\gamma}{\sqrt{{\cal S}^2+{\cal P}^2}}\Bigr)F-\frac{{\cal P}\sinh\gamma}{\sqrt{{\cal S}^2+{\cal P}^2}}\Tilde{F}\,.
\end{equation}

In the limit $\mathcal{S}\to 0$, ModMax electrodynamics smoothly reduces to linear Maxwell theory, so that purely electric or purely magnetic static configurations coincide with their Maxwellian counterparts (up to trivial $\gamma$-dependent normalisations). When both electric and magnetic charges are present, spherically symmetric static solutions show only perturbative deviations from the Einstein-Maxwell profiles, with these deviations governed by the ModMax deformation parameter $\gamma$. By contrast, the inclusion of rotation (even at the level of a slow-rotation expansion) activates the theory's intrinsic nonlinearity and modifies the spacetime and field structure in a nontrivial way. Consequently, despite advancements made through perturbative and numerical techniques, the literature to date lacks any exact closed-form solution that describes a fully rotating, charged ModMax spacetime.

Drawing on this context, an exact analytical solution for black holes can be derived if the following assumptions are taken into account: (i) the charged configuration is supported only by the interaction Lagrangian; and (ii) the interaction is implemented by modulating the Kalb-Ramond three-form field strength $H_{\mu\nu\rho}$ through the addition of a Chern-Simons triple form constructed from the $U(1)$ gauge potential. Introducing the gauge-Kalb-Ramond coupling through the generalized field strength
\begin{equation}
\tilde H_{\mu\nu\rho}=H_{\mu\nu\rho}+A_{[\mu}P_{\nu\rho]},
\end{equation}
which couples the Kalb-Ramond sector to the Lorentz electromagnetic invariants, one finds that the mixed and quadratic interaction contributions vanish identically in the modified kinetic term.

Under the purely electric ansatz adopted above, all interaction contributions in the modified kinetic term vanish identically:
\begin{equation}
\tilde{H}^{\mu\nu\rho}\tilde{H}_{\mu\nu\rho}
=H^{\mu\nu\rho}H_{\mu\nu\rho}+2H^{\mu\nu\rho}A_{[\mu}P_{\nu\rho]}+A^{[\mu}P^{\nu\rho]}A_{[\mu}P_{\nu\rho]}=0 .
\end{equation}

Consequently, the matter sector is governed solely by the electromagnetic invariants together with a nontrivial nonminimal interaction factor. The matter Lagrangian may be written in the compact form
\begin{equation}
\label{MT}
\mathcal{L}_{\mathrm{M}}
= -\,\mathcal{S}\bigl(1+2\eta B^{\alpha\beta}B_{\alpha\beta}\bigr)\cosh\gamma
+ \sqrt{\mathcal{S}^{2}+\mathcal{P}^{2}}\bigl(1+2\eta B^{\alpha\beta}B_{\alpha\beta}\bigr)\sinh\gamma,
\end{equation}
where $\eta$ denotes the nonminimal coupling constant. Under the $SO(2)$ duality rotations the factor $(1+2\eta B^{2})$ transforms as a scalar and therefore does not spoil the duality structure of the ModMax sector.

If the Kalb-Ramond two-form acquires a nonzero vacuum expectation value, the theory undergoes spontaneous local Lorentz-symmetry breaking (LSB) in the electromagnetic/duality sector. In that case the effective electromagnetic couplings are modified by the background $B_{\mu\nu}$ and the nonminimal factor $(1+2\eta B^{2})$, and the altered coupling structure permits the existence of charged black hole solutions in the coupled KR-ModMax system.

Varying the action \eqref{Main_Action} with respect to the inverse metric $g^{\mu\nu}$ yields the modified Einstein equations which couple gravity to both the matter (electromagnetic) sector and the KR field,
\begin{equation}
\label{EoM1}
R_{\mu\nu}-\tfrac{1}{2}g_{\mu\nu}R = T^{\mathrm{matter}}_{\mu\nu} + T^{\mathrm{KR}}_{\mu\nu},
\end{equation}
where $T^{\mathrm{matter}}_{\mu\nu}$ denotes the stress-energy tensor of the electromagnetic (ModMax) sector and $T^{\mathrm{KR}}_{\mu\nu}$ is the effective stress-energy tensor of the KR field.

An explicit expression for the matter contribution reads
\begin{align}
\label{T_matter_rephrased}
T_{\mu\nu}^{\mathrm{matter}} &= \bigl(1 + 2\eta\, B^{\rho\sigma}B_{\rho\sigma}\bigr)\Bigg\{
F_{\mu\lambda}F_{\nu}{}^{\lambda}\!\left(-\cosh\gamma + \frac{\mathcal{S}}{\sqrt{\mathcal{S}^2+\mathcal{P}^2}}\sinh\gamma\right)\nonumber\\[4pt]
&\qquad\qquad\qquad
+\,\tilde{F}_{\mu\lambda}\tilde{F}_{\nu}{}^{\lambda}\!\left(\frac{\mathcal{P}}{\sqrt{\mathcal{S}^2+\mathcal{P}^2}}\sinh\gamma\right)
+ g_{\mu\nu}\!\left(-\mathcal{S}\cosh\gamma + \sqrt{\mathcal{S}^2+\mathcal{P}^2}\sinh\gamma\right)\Bigg\}\nonumber\\[4pt]
&\qquad\qquad
-\,4\eta\!\left(-\mathcal{S}\cosh\gamma + \sqrt{\mathcal{S}^2+\mathcal{P}^2}\sinh\gamma\right)B_{\mu\lambda}B_{\nu}{}^{\lambda}.
\end{align}

The KR-sector contributes the following effective energy-momentum tensor:
\begin{align}
\label{rr}
T^{\mathrm{KR}}_{\mu\nu} &= \tfrac{1}{2}H_{\mu\alpha\beta}H_{\nu}{}^{\alpha\beta}
- \tfrac{1}{12} g_{\mu\nu} H^{\alpha\beta\rho}H_{\alpha\beta\rho}
+ 2V' B_{\alpha\mu}B^{\alpha}{}_{\nu} - g_{\mu\nu}V \nonumber\\[4pt]
&\quad + \xi_{2}\Bigg\{\tfrac{1}{2} g_{\mu\nu}\bigl(B^{\alpha\gamma}B^{\beta}{}_{\gamma}R_{\alpha\beta}\bigr)
- \bigl(B^{\alpha}{}_{\mu}B^{\beta}{}_{\nu}R_{\alpha\beta}\bigr)
- \bigl(B^{\alpha\beta}B_{\nu\beta}R_{\mu\alpha}\bigr)
- \bigl(B^{\alpha\beta}B_{\mu\beta}R_{\nu\alpha}\bigr)\Bigg\}\nonumber\\[4pt]
&\quad + \tfrac{1}{2}\xi_{2}\Bigl(\nabla_{\alpha}\nabla_{\mu}\!\bigl(B^{\alpha\beta}B_{\nu\beta}\bigr)
+ \nabla_{\alpha}\nabla_{\nu}\!\bigl(B^{\alpha\beta}B_{\mu\beta}\bigr)\Bigr)\nonumber\\[4pt]
&\quad - \tfrac{1}{2}\xi_{2}\Bigl(\nabla^{\alpha}\nabla_{\alpha}\!\bigl(B_{\mu}{}^{\gamma}B_{\nu\gamma}\bigr)
+ g_{\mu\nu}\nabla_{\alpha}\nabla_{\beta}\!\bigl(B^{\alpha\gamma}B^{\beta}{}_{\gamma}\bigr)\Bigr).
\end{align}

Throughout, primes denote derivatives with respect to the function's argument. Finally, the total stress-energy tensor is covariantly conserved,
\begin{equation}
\nabla^{\mu}\bigl(T^{\mathrm{KR}}_{\mu\nu}+T^{\mathrm{matter}}_{\mu\nu}\bigr)=0,
\end{equation}
which follows from diffeomorphism invariance of the action and ensures consistency of the coupled system.

Varying the action \eqref{Main_Action} with respect to the Kalb--Ramond two-form \(B^{\mu\nu}\) yields the KR field equation of motion,
\begin{align}
\label{KK}
\nabla^{\alpha}H_{\alpha\mu\nu}
+3\xi_{2}\,R_{\alpha[\mu}B^{\alpha}{}_{\nu]}
-6V'\bigl(B^{2}\bigr)\,B_{\mu\nu}
-4\eta\,\Bigl(-\mathcal{S}\cosh\gamma+\sqrt{\mathcal{S}^{2}+\mathcal{P}^{2}}\,\sinh\gamma\Bigr)\,B_{\mu\nu}\qquad\qquad\qquad\qquad\qquad \nonumber\\
-\,4\eta\Bigl(\mathcal{L}_{\mathcal{S}}F^{\rho\sigma}+\mathcal{L}_{\mathcal{P}}\tilde{F}^{\rho\sigma}\Bigr)F_{\mu\nu}B_{\rho\sigma}=0,
\end{align}
where we introduced the short-hand derivative coefficients
\begin{equation}
\mathcal{L}_{\mathcal{S}} \equiv -\cosh\gamma + \frac{\mathcal{S}}{\sqrt{\mathcal{S}^{2}+\mathcal{P}^{2}}}\sinh\gamma,
\qquad
\mathcal{L}_{\mathcal{P}} \equiv \frac{\mathcal{P}}{\sqrt{\mathcal{S}^{2}+\mathcal{P}^{2}}}\sinh\gamma .
\end{equation}

Likewise, variation of \eqref{Main_Action} with respect to the gauge potential $A^{\mu}$ produces the field equations of ModMax electrodynamics in the presence of KR coupling:
\begin{equation}
\label{MDM}
\nabla^{\nu}\Biggl\{\Bigl[\mathcal{L}_{\mathcal{S}}\,F_{\mu\nu}
+\mathcal{L}_{\mathcal{P}}\,\tilde{F}_{\mu\nu}\Bigr]\bigl(1+2\eta\,B^{\alpha\beta}B_{\alpha\beta}\bigr)\Biggr\}=0.
\end{equation}
In the limit $\eta\to0$ these equations reduce to the standard ModMax nonlinear Maxwell equations.

To construct explicit black-hole solutions we concentrate on the electrically charged sector and on static, spherically symmetric configurations. In particular we restrict to the purely electric branch by imposing $\mathcal{P}=0$ and adopting the electrostatic ansatz $A_{\mu}=\Phi(r)\,\delta^{0}_{\mu}$.

Under these assumptions the KR vacuum is characterized by a single nonvanishing component,
\begin{equation}
\label{KR_VEV_rephrased}
b_{10}=-b_{01}=\tilde{E}(r),
\end{equation}
and all remaining background components vanish. Consequently the KR field strength identically vanishes on this background,
\begin{equation}
H_{\lambda\mu\nu}=0,
\end{equation}
so that the KR dynamics on the chosen vacuum are governed solely by the algebraic potential terms, curvature couplings and the electromagnetic backreaction encoded in \eqref{KK} and \eqref{MDM}.

Finally, we remark that restricting to $\mathcal{P}=0$ and $A_{\mu}=\Phi(r)\delta^{0}_{\mu}$ isolates the interplay between the triplet-form $U(1)$ Chern-Simons-type electromagnetic structure and the ModMax sector, while keeping the analysis tractable for the electrically charged black hole ansatz considered below.

\section{Electrically charged black hole solutions}\label{sec:Solution}
To solve the field equations (\ref{EoM1})--(\ref{rr}) we adopt a static, spherically symmetric four-dimensional line element given by the metric ansatz
\begin{equation}
\label{Metric}
ds^{2}=-\mathbb{F}_{1}(r)\,dt^{2}+\mathbb{F}_{2}(r)\,dr^{2}+r^{2}d\theta^{2}+r^{2}\sin^{2}\theta\,d\phi^{2},
\end{equation}
where \(\mathbb{F}_{1}(r)\) and \(\mathbb{F}_{2}(r)\) are radial metric functions to be determined from the field equations. Substituting this ansatz into Eqs.~(\ref{EoM1})--(\ref{rr}) the Kalb--Ramond vacuum profile may be written as
\begin{equation}
\label{KR_VEV_redef}
\tilde{E}(r)=|b|\sqrt{\tfrac{1}{2}\,\mathbb{F}_{1}(r)\,\mathbb{F}_{2}(r)} .
\end{equation}
Accordingly, the vacuum expectation value of the KR two-form satisfies the fixed-norm constraint
\begin{equation}
b^{\mu\nu}b_{\mu\nu}=-b^{2}.
\end{equation}

Consequently, substituting the Kalb--Ramond vacuum configuration into the gravitational field equations (\ref{EoM1}) yields the modified Einstein equations in the compact form
\begin{align}
\label{EoM2}
R_{\mu\nu} &= T^{\mathrm{M}}_{\mu\nu} - \tfrac{1}{2}\,g_{\mu\nu}T^{\mathrm{M}}
+ V'\!\left(2\,b_{\mu\alpha}b_{\nu}{}^{\alpha} + b^{2}g_{\mu\nu}\right) \nonumber\\
&\qquad + \xi_{2}\Bigl\{\,g_{\mu\nu}\,b^{\alpha\gamma}b^{\beta}{}_{\gamma}\,R_{\alpha\beta}
- b^{\alpha}{}_{\mu}b^{\beta}{}_{\nu}\,R_{\alpha\beta}
- b^{\alpha\beta}b_{\mu\beta}\,R_{\nu\alpha}
- b^{\alpha\beta}b_{\nu\beta}\,R_{\mu\alpha} \nonumber\\
&\qquad\qquad\; + \tfrac{1}{2}\nabla_{\alpha}\nabla_{\mu}\!\left(b^{\alpha\beta}b_{\nu\beta}\right)
+ \tfrac{1}{2}\nabla_{\alpha}\nabla_{\nu}\!\left(b^{\alpha\beta}b_{\mu\beta}\right)
- \tfrac{1}{2}\nabla^{\alpha}\nabla_{\alpha}\!\left(b_{\mu}{}^{\gamma}b_{\nu\gamma}\right)\Bigr\}\,.
\end{align}

Here $T^{\mathrm{M}} \equiv g^{\alpha\beta}T^{\mathrm{M}}_{\alpha\beta}$ is the trace of the matter stress-energy tensor, and $V'$ denotes the derivative of the KR self-interaction potential evaluated on the vacuum configuration. For electrically ModMax electrodynamics one has the traceless property $T^{\mu}{}_{\mu}=0$.

At this stage we concentrate on solving the modified field equations \eqref{EoM2}.  We adopt the static, spherically symmetric metric ansatz \eqref{Metric}, assume a purely electrostatic gauge potential $A_{\mu}=\Phi(r)\,\delta^{t}_{\mu}$, and redefine
$\ell \equiv \xi_{2}b^{2}/2$, so that $\ell$ quantifies the amplitude of Lorentz-symmetry breaking induced by the Kalb-Ramond vacuum. With these assumptions the coupled gravitational-KR-electromagnetic system reduces to the following set of ordinary differential equations (primes denote $d/dr$).

\begin{subequations}\label{EoMs}
\begin{align}
&\frac{2\mathbb{F}_{1}''}{\mathbb{F}_{1}}
-\frac{\mathbb{F}_{1}'}{\mathbb{F}_{1}}\frac{\mathbb{F}_{2}'}{\mathbb{F}_{2}}
-\left(\frac{\mathbb{F}_{1}'}{\mathbb{F}_{1}}\right)^{2}
+\frac{4}{r}\frac{\mathbb{F}_{1}'}{\mathbb{F}_{1}}
-4\frac{1-2\eta b^{2}}{(1-\ell)\,\mathbb{F}_{1}}\bigl(\cosh\gamma+\sinh\gamma\bigr)\,(\Phi')^{2}
=0, \label{EoM_1} \\[6pt]
&\frac{2\mathbb{F}_{1}''}{\mathbb{F}_{1}}
-\frac{\mathbb{F}_{1}'}{\mathbb{F}_{1}}\frac{\mathbb{F}_{2}'}{\mathbb{F}_{2}}
-\left(\frac{\mathbb{F}_{1}'}{\mathbb{F}_{1}}\right)^{2}
-\frac{4}{r}\frac{\mathbb{F}_{2}'}{\mathbb{F}_{2}}
-4\frac{1-2\eta b^{2}}{(1-\ell)\,\mathbb{F}_{1}}\bigl(\cosh\gamma+\sinh\gamma\bigr)\,(\Phi')^{2}
=0, \label{EoM_2} \\[6pt]
&\frac{2\mathbb{F}_{1}''}{\mathbb{F}_{1}}
-\frac{\mathbb{F}_{1}'\mathbb{F}_{2}'}{\mathbb{F}_{1}\mathbb{F}_{2}}
-\left(\frac{\mathbb{F}_{1}'}{\mathbb{F}_{1}}\right)^{2}
+\frac{1+\ell}{\ell\,r}\!\left(\frac{\mathbb{F}_{1}'}{\mathbb{F}_{1}}-\frac{\mathbb{F}_{2}'}{\mathbb{F}_{2}}\right)
+\frac{2(1-\ell)}{\ell\,r^{2}}
-\frac{2\mathbb{F}_{2}}{\ell\,r^{2}}\bigl(1-b^{2}r^{2}V'\bigr) \nonumber\\[4pt]
&\qquad\qquad\qquad
-2\frac{1-6\eta b^{2}}{\ell\,\mathbb{F}_{1}}\bigl(\cosh\gamma+\sinh\gamma\bigr)\,(\Phi')^{2}
=0. \label{EoM_3}
\end{align}
\end{subequations}

The Kalb-Ramond field equation \eqref{KK} and the modified Maxwell equation \eqref{MDM} take the explicit form
\begin{align}
&\frac{2\mathbb{F}_{1}''}{\mathbb{F}_{1}}
-\left(\frac{\mathbb{F}_{1}'}{\mathbb{F}_{1}}\right)^{2}
+\frac{2}{r}\!\left(\frac{\mathbb{F}_{1}'}{\mathbb{F}_{1}}-\frac{\mathbb{F}_{2}'}{\mathbb{F}_{2}}\right)
-\frac{\mathbb{F}_{1}'}{\mathbb{F}_{1}}\frac{\mathbb{F}_{2}'}{\mathbb{F}_{2}}
+\frac{2b^{2}V'\,\mathbb{F}_{2}}{\ell}
- \frac{4\eta b^{2}}{\ell\,\mathbb{F}_{1}}\bigl(\cosh\gamma+\sinh\gamma\bigr)\,(\Phi')^{2}
=0, \label{EOM_KR_2} \\[6pt]
&\bigl(\cosh\gamma+\sinh\gamma\bigr)\!\left[\Phi''+\frac{\Phi'}{2}\!\left(\frac{4}{r}-\frac{\mathbb{F}_{1}'}{\mathbb{F}_{1}}-\frac{\mathbb{F}_{2}'}{\mathbb{F}_{2}}\right)\right]\!(1-2\eta b^{2})=0.
\label{Maxwell_EQ_2}
\end{align}

These equations form the starting point for the integration strategy: Eqs.~\eqref{EoM_1}--\eqref{EoM_3} encode the modified gravitational dynamics while \eqref{EOM_KR_2} and \eqref{Maxwell_EQ_2} govern the KR profile and the electrostatic potential respectively. Subsequent simplifications may be obtained by fixing the potential $V$, choosing a gauge for $\Phi$, or imposing algebraic relations between $\mathbb{F}_{1}$ and $\mathbb{F}_{2}$ (e.g. $\mathbb{F}_{2}=1/\mathbb{F}_{1}$).

In this work we neglect the cosmological constant and assume that the Kalb--Ramond vacuum sits at a stationary point of the self-interaction potential, i.e. \(V' = 0\). Accordingly we take the simplest nontrivial self-interaction,
\begin{equation}
V(X)=\tfrac{1}{2}\,\lambda\,X^{2},\qquad
X\equiv B^{\mu\nu}B_{\mu\nu}+b^{2},
\end{equation}
where \(\lambda\) is the coupling constant and the vacuum is located at \(X=0\).

Subtracting Eq.~\eqref{EoM_2} from Eq.~\eqref{EoM_1} yields the algebraic relation
\begin{equation}
\frac{\mathbb{F}_1'}{\mathbb{F}_2'}+\frac{\mathbb{F}_1}{\mathbb{F}_2}=0,
\end{equation}
which is equivalent to
\begin{equation}
\bigl(\mathbb{F}_1\mathbb{F}_2\bigr)'=0.
\end{equation}
Hence the product $\mathbb{F}_1(r)\mathbb{F}_2(r)$ is constant:
\begin{equation}
\mathbb{F}_1(r)\,\mathbb{F}_2(r)=C.
\label{Relation_dFdG_general}
\end{equation}
The constant $C$ may be absorbed by a redefinition (rescaling) of the time coordinate; therefore, without loss of generality we set $C=1$ and adopt the convenient gauge
\begin{equation}
\mathbb{F}_2(r)=\mathbb{F}_1^{-1}(r),
\label{Relation_FG}
\end{equation}
which brings the line element \eqref{Metric} into the familiar Schwarzschild-like form and simplifies the ensuing integration of the field equations.

Imposing the metric and electrostatic  ansatze into the modified Maxwell equation \eqref{Maxwell_EQ_2} yields the ordinary differential equation
\begin{equation}
\Phi''+\frac{2}{r}\Phi'=0,
\end{equation}
whose general solution is
\begin{equation}
\Phi(r)=\frac{\mathcal{A}_1}{r}+\Phi_0,
\end{equation}
with $\mathcal{A}_1$ and \(\Phi_0\) integration constants. Choosing the gauge \(\Phi(\infty)=0\) sets \(\Phi_0=0\), so that
\begin{equation}
\Phi(r)=\frac{\mathcal{A}_1}{r}.
\end{equation}

The conserved current is modified to
\begin{equation}
J^{\mu}=\nabla_{\nu}\!\Bigl(P^{\mu\nu}+2\eta\,B^{\mu\nu}B^{\alpha\beta}P_{\alpha\beta}\Bigr),
\end{equation}
and the total electric charge \(Q\) is obtained from the surface integral at spatial infinity via Stokes' theorem:
\begin{align}
Q &= -\frac{1}{4\pi}\int_{S^2_{\infty}} d^{3}x\sqrt{\gamma^{(3)}}\,n_{\mu}J^{\mu}
= -\frac{1}{4\pi}\int_{\partial S^2_{\infty}} d\theta d\phi\sqrt{\gamma^{(2)}}\,n_{\mu}\sigma_{\nu}
\Bigl(P^{\mu\nu}+2\eta\,B^{\mu\nu}B^{\alpha\beta}P_{\alpha\beta}\Bigr) \nonumber\\
&= \bigl(\cosh\gamma+\sinh\gamma\bigr)\bigl(1-2\eta b^{2}\bigr)\,\mathcal{A}_1,
\end{align}
where $S^{2}_{\infty}$ denotes a spacelike slice with induced metric $\gamma^{(3)}_{ij}$, $n_{\mu}=(1,0,0,0)$ is its unit normal, \(\partial S^{2}_{\infty}\) is the two-sphere at infinity with metric $\gamma^{(2)}_{ij}=r^{2}(d\theta^{2}+\sin^{2}\theta\,d\phi^{2})$, and $\sigma_{\mu}=(0,1,0,0)$ is the outward radial unit normal on the two-sphere.

Solving for $\mathcal{A}_1$ in terms of the physical charge $Q$ gives
\begin{equation}
\mathcal{A}_1=\frac{Q}{\bigl(\cosh\gamma+\sinh\gamma\bigr)\bigl(1-2\eta b^{2}\bigr)}.
\end{equation}
Noting $\cosh\gamma+\sinh\gamma=e^{\gamma}$, the electrostatic potential therefore reads
\begin{equation}\label{Phi}
\Phi(r)
=\frac{Q\,e^{-\gamma}}{\bigl(1-2\eta b^{2}\bigr)\,r},
\end{equation}
which reduces to the standard Coulomb potential in the limit $\gamma\to0$ and $\eta\to0$.

Proceeding from the previous results, subtracting Eq.~\eqref{EoM_3} from Eq.~\eqref{EoM_1} and inserting the gauge choice \eqref{Relation_FG} together with the electrostatic profile \eqref{Phi} leads to the ordinary differential equation
\begin{equation}
\label{ode_for_F1}
-\frac{2(1-\ell)}{\ell\,r}\,\frac{\mathbb{F}_1'(r)}{\mathbb{F}_1(r)}
+\frac{2}{\ell\,r^2\,\mathbb{F}_1(r)}
-\frac{2(1-\ell)}{\ell\,r^2}
-2\,\frac{e^{-\gamma}Q^2\bigl[\,1+\ell-2(3-\ell)\eta b^2\bigr]}
{\ell\,(1-\ell)(1-2\eta b^2)^2\,\mathbb{F}_1(r)\,r^4}
=0.
\end{equation}

Integrating \eqref{ode_for_F1} yields an exact, closed-form expression for the metric function \(\mathbb{F}_1(r)\):
\begin{equation}
\label{F1_general}
\mathbb{F}_1(r)
=\frac{1}{1-2\eta b^2}+\frac{C_1}{r}
+\frac{e^{-\gamma}Q^2\bigl[\,1+\ell-2(3-\ell)\eta b^2\bigr]}
{(1-\ell)^2(1-2\eta b^2)^2\,r^2}\,,
\end{equation}
where $C_1$ is an integration constant. Demanding that the solution recover the Schwarzschild limit when the additional parameters are switched off $(Q\to0$, \(\ell\to0\), \(\eta\to0\)) fixes the integration constant to \(C_1=-2M\), so that the geometry is asymptotically Schwarzschild-like.

A consistency check of the full system (Eqs.~\eqref{EoM_1}--\eqref{EoM_3}, \eqref{EOM_KR_2}, \eqref{Maxwell_EQ_2}) shows that the coupled field equations admit physically acceptable, charged black hole solutions only if the interaction coupling satisfies the relation
\begin{equation}
\eta=\frac{\ell}{2b^2}.
\label{eta_constraint}
\end{equation}
Thus the interaction Lagrangian $\mathcal{L}_{\mathrm{int}}$ plays an essential role: in the presence of Lorentz-symmetry breaking (nonzero $\ell$) the interaction sector is a necessary ingredient for the existence of a charged static, spherically symmetric solution.

Under the constraint \eqref{eta_constraint} the electrostatic potential and the metric function simplify to the compact forms
\begin{align}
\label{Phi_final}
\Phi(r)&=\frac{e^{-\gamma} Q}{\left(1-\ell \right)r},\\[6pt]
\mathbb{F}_1(r)&=\frac{1}{1-\ell}-\frac{2M}{r}+\frac{e^{-\gamma }   Q^2}{(1-\ell)^2 r^2}.\label{Solution_Fr_RN}
\end{align}
Equation \eqref{Solution_Fr_RN} is manifestly a Reissner-Nordstrom-type profile dressed by the Lorentz-violating parameter $\ell$ and the ModMax parameter $\gamma$; in the limit $\ell\to0$ the solution reduces to the static, spherically symmetric charged solution previously obtained for ModMax electrodynamics~\cite{Flores2021}.  

\noindent Remarkably, the combination of the rescaling factor $1/(1-\ell)$ and the exponential factor $e^{-\gamma}$ encodes the two distinct deformations of the standard RN geometry: the former originates from Lorentz-symmetry breaking induced by the KR vacuum, while the latter stems from the nonlinear ModMax electromagnetic sector.

%%%%%%%%%%%%%%%%%%%%%%%%%%%%%%%%%%%%%%%%%%%%%%%%%%%%%%%%%%%%%%%%

%%%%%%%%%%%%%%%%%%%%%%%%%%%%%%%%%%%%%%%%%%%%%%%%%%%%%%%%%%%%%%%%%%

\begin{figure*}[hbt!]
\begin{tabular}{c c}
\includegraphics[scale=0.65]{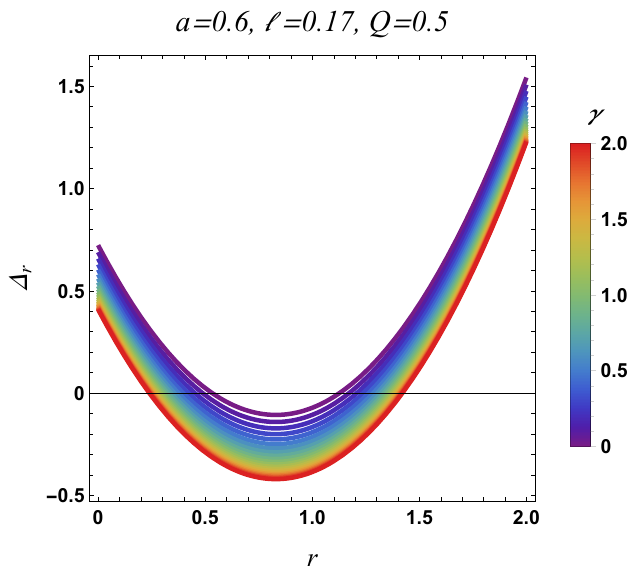}
\includegraphics[scale=0.65]{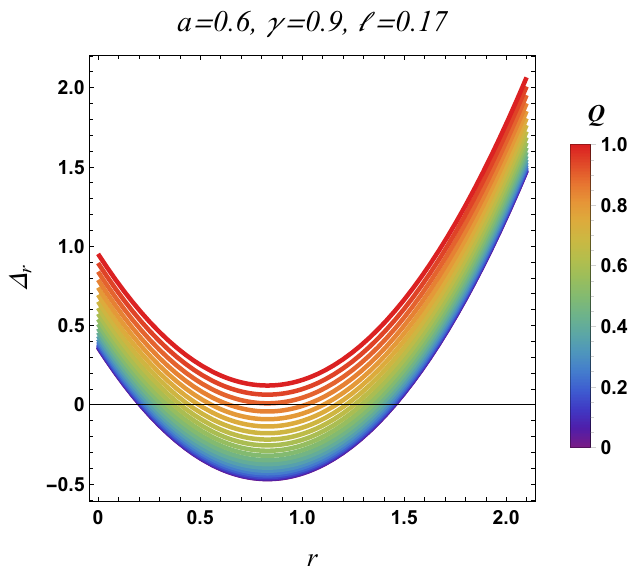}\\
\includegraphics[scale=0.65]{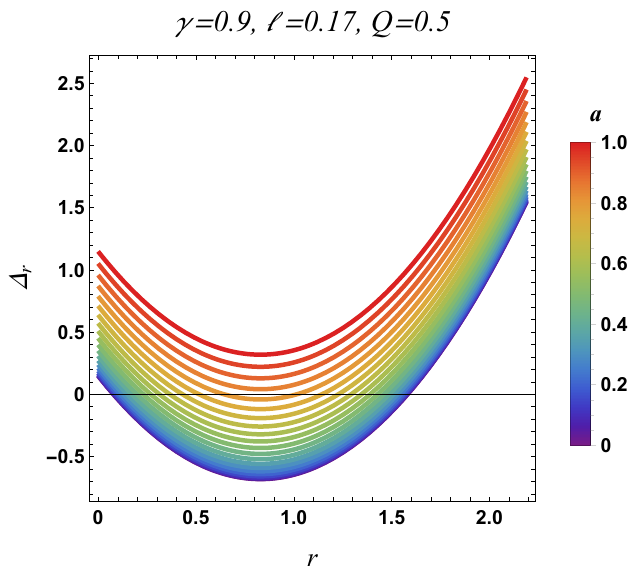}
\includegraphics[scale=0.65]{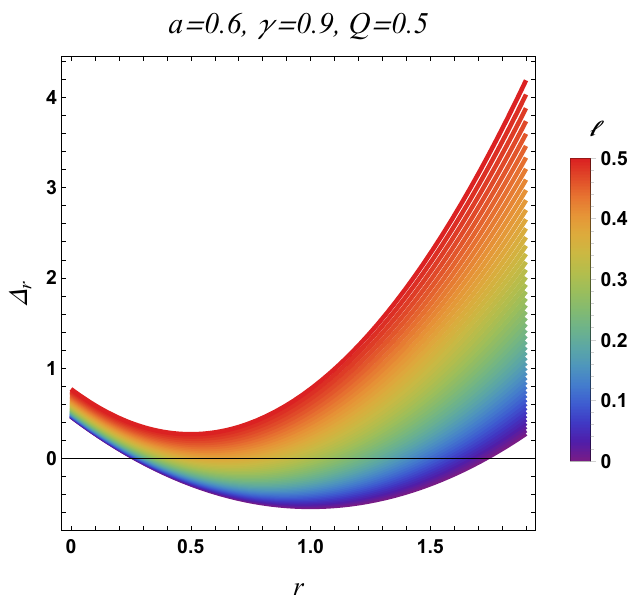}
\end{tabular}	
\caption{Radial dependence of the horizon function $\Delta_{r}$ as a function of the coordinate radius $r$, 
shown for representative values of the parameter space $(\gamma, Q, a, \ell)$.}
\label{FigM87}
\end{figure*}

\section{Rotating KR BHs coupled to Traceless Conformal Electrodynamics}\label{sec:MNJ}
In this section we introduce a spin parameter into our static solution \eqref{Solution_Fr_RN} (with $\mathbb{F}_2(r)=\mathbb{F}_1(r)^{-1}$) by applying the original Newman-Janis algorithm (NJA) \cite{Newman:1965tw}. Unlike the modified Newman-Janis algorithm (MNJA) that replaces the final step by a non-complexification of the radial coordinate \cite{Azreg-Ainou:2014pra,Azreg-Ainou:2014aqa,Azreg-Ainou:regular}. In the BH literature, numerous studies have investigated the MNJA in GR and in modified gravity theories (MOG)~\cite{Azreg-Ainou:2014aqa, Johannsen:2011dh, Ghosh:2014pba, Kumar:2017qws, Kumar:2019pjp, Kumar:2020hgm, Kumar:2020owy, Ghosh:2021clx, Ghosh:2023nkr}; it has also been used to construct rotating BHs in Kalb-Ramond gravity~\cite{Brahma:2020eos, Kumar:2022vfg}. Here we follow the standard NJA complexification procedure and construct the rotating configuration. Hence, this method enables us to derive the appropriate rotating metric in Boyer-Lindquist coordinates, expressed by
\begin{align}
ds^2 =&\frac{\Psi(r,\theta,a)}{\rho^2}\Big[ - \Big(1-\frac{2\mathcal{F}}{\rho^2}\Big) dt^2 - 2\frac{2\mathcal{F}}{\rho^2}a\sin^2\theta dt d\phi 
+ \frac{\Sigma}{\rho^2}\sin^2\theta d\phi^2 + \frac{\rho^2}{\Delta_r}dr^2 + \rho^2d\theta^2\Big]\nonumber\\
\label{xian0}=&\frac{\Psi}{\rho^2}\Big[ - \Big(\frac{\Delta_r-a^2\sin^2\theta}{\rho^2}\Big) dt^2 +2a\sin^2\theta\Big(\frac{\Delta_r-a^2-r^2}{\rho^2}\Big)d td\phi 
+ \frac{(r^2+a^2)^2 - a^2\Delta_r\sin^2\theta}{\rho^2}\sin^2\theta d\phi^2 + \frac{\rho^2}{\Delta_r}dr^2 + \rho^2d\theta^2\Big],
\end{align}
where
\begin{equation}
\begin{split}
\rho^2(r,\theta) &\equiv r^2 + a^2 \cos^2\theta,\qquad 2\mathcal{F}(r)\equiv r^2(1-F)=r^2\Big(1-\frac{1}{1-\ell}+\frac{2M}{r}-\frac{e^{-\gamma }   Q^2}{(1-\ell)^2 r^2}\Big)\\
\Delta_r(r)&\equiv r^2F + a^2 = \frac{r^2}{1-\ell}-2M\,r+\frac{e^{-\gamma }   Q^2}{(1-\ell)^2}+a^{2}\\
\Sigma(r,\theta)&\equiv (r^2+a^2)^2 - a^2\Delta_r\sin^2\theta .
\end{split}\label{eq:ds_kerr_like}
\end{equation}
Here $a = J/M$ is spin, where $J$ is the BH's angular momentum and $M$ the ADM mass. The function $\Psi(r,\theta,a)$ satisfies the nonlinear differential equation $G_{r\theta}=0$ ($G_{\mu\nu}$ is the Eistein tensor)~\cite{Azreg-Ainou:2014pra}:
\begin{equation}
(r^2+a^2 z)^2(3\partial_r\Psi\partial_{z}\Psi-2\Psi\partial_{rz}\Psi) = 3a^2\partial_r r^2\Psi^2,
\label{eq:Psi_cond_1}
\end{equation}
where $z \equiv \cos^2\theta$ and is given by $\Psi =\rho^2=r^2+a^2 z=r^2 + a^2 \cos^2\theta$~\cite{Azreg-Ainou:2014pra} and the rotating metric takes~\eqref{xian0} the simple form:
\begin{equation}\label{xian}
ds^2 = - \frac{\Delta_r-a^2\sin^2\theta}{\rho^2} dt^2 +2a\sin^2\theta\frac{\Delta_r-a^2-r^2}{\rho^2}d td\phi 
+ \frac{(r^2+a^2)^2 - a^2\Delta_r\sin^2\theta}{\rho^2}\sin^2\theta d\phi^2 + \frac{\rho^2}{\Delta_r}dr^2 + \rho^2d\theta^2.
\end{equation}% 
%The above condition ensures the vanishing of the $r\theta$-component of the Einstein tensor for an imperfect fluid rotating about the $z$-axis. For definiteness, the metric (\ref{xian}) is termed the RQBH. The Kerr black hole is recovered by setting $b_0 = 1$ and $Q = 0$ in the metric functions $A(r)$ and $B(r)$. However, since $A(r) \neq B(r)$ in \eqref{eq:ds_0}, fixing $\Psi$ is nontrivial. When $a = 0$, metric \eqref{xian} reduces to the static form \eqref{eq:ds_0} for $\Psi = r^2$. Notably,  $\Psi$ acts as a conformal factor in \eqref{xian} and does not affect the causal structure or null geodesics in the exterior. As $b_0 \to 1$ and $Q = 0$, \eqref{xian} recovers the Kerr metric, and in the limit $a \to 0$, it reduces to Schwarzschild. We refer to metric \eqref{xian} as the RQBH.
%We keep $\Psi$ unspecified, and our results hold independently of its form.

The metric (\ref{xian}) exhibits a coordinate singularity at points where $\Sigma \neq 0$ and $\Delta_r = 0$, which defines the event horizon (EH). Figure~\ref{FigM87} represents $\Delta_{r}(r)$ as the main diagnostic of the existence and multiplicity of the horizon. The analytical structure of $\Delta_{r}$ reveals that the ModMax non-linearity $\gamma$ appears in the metric as a multiplicative rescaling of the Maxwell contribution (via $e^{-\gamma} Q^{2}$), while the Kalb-Ramond-induced amplitude $\ell$, which violates Lorentz symmetry, modifies the global radial coefficients by the factors $(1-\ell)^{-1}$ and $(1-\ell)^{-2}$~\eqref{eq:ds_kerr_like}. Consequently, increasing $\gamma$ (for fixed $Q$) reduces the effective electromagnetic interaction on the geometry and tends to shift the outer horizon $r_{+}$ outward compared to the underlying case $\gamma=0$; conversely, increasing $Q$ or $\ell$ tends to drag $r_{+}$ inwards and can lead the system towards extremality or naked singularity branches. These numerical sections therefore highlight the rival roles of $\gamma$ (charge screen) and $\ell$ (metric rescaling) in determining the structure of the horizon.

\begin{figure}[hbt!]
  	\centering
\includegraphics[scale=0.45]{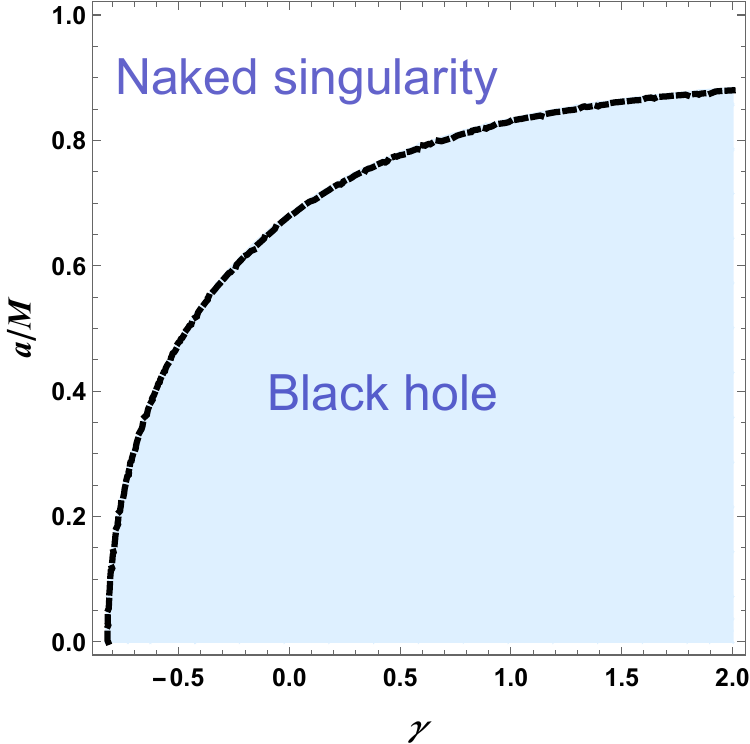}~~~
\includegraphics[scale=0.45]{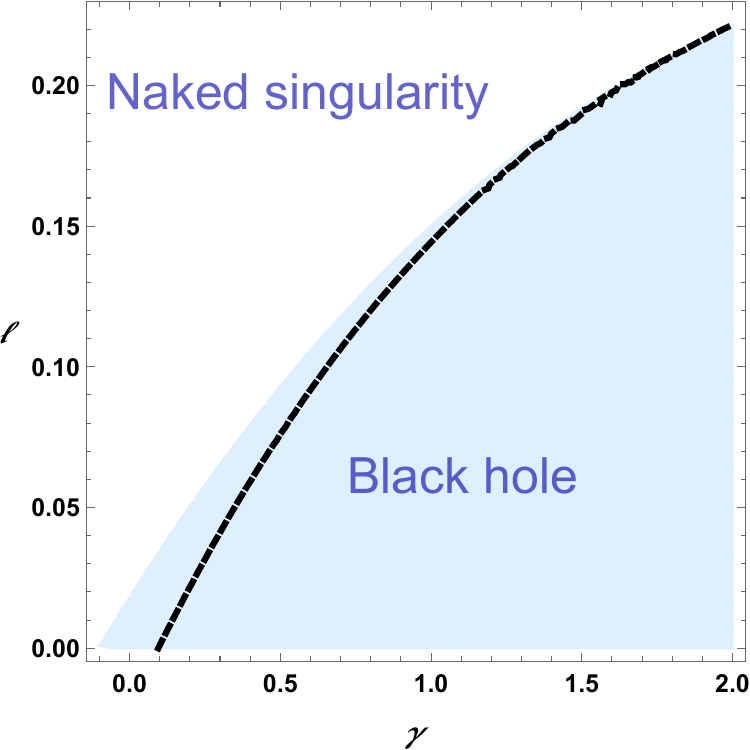}~~~~
\includegraphics[scale=0.45]{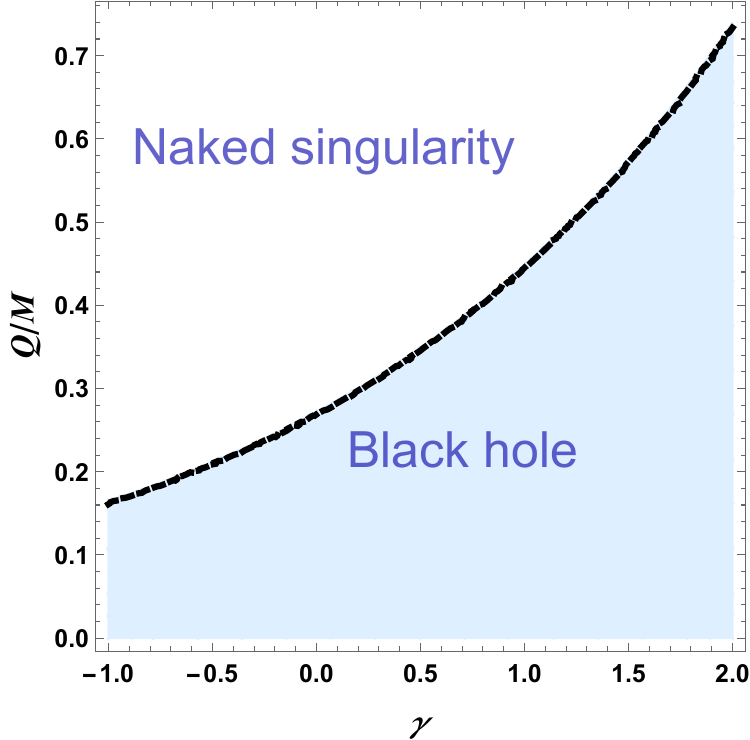}~~~
\caption{The parameter spaces $(\gamma,a/M)$, $(\gamma,\ell)$ and $(\gamma,Q/M)$ for the RKRBH metric. The parameters on the black curve correspond to extremal rotating BHs with degenerate horizons. The blue curve separates the BH from naked singularity.} \label{Figps}
\end{figure}

\begin{figure}[hbt!]
  	\centering    \includegraphics[scale=0.78]{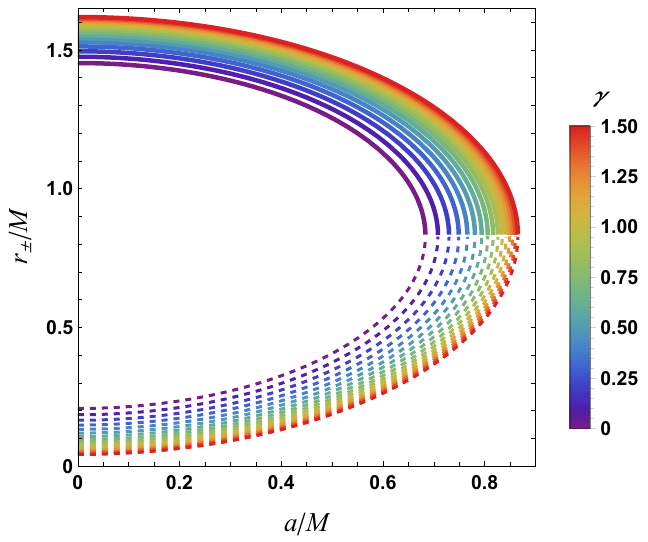}~~~
  	\includegraphics[scale=0.78]{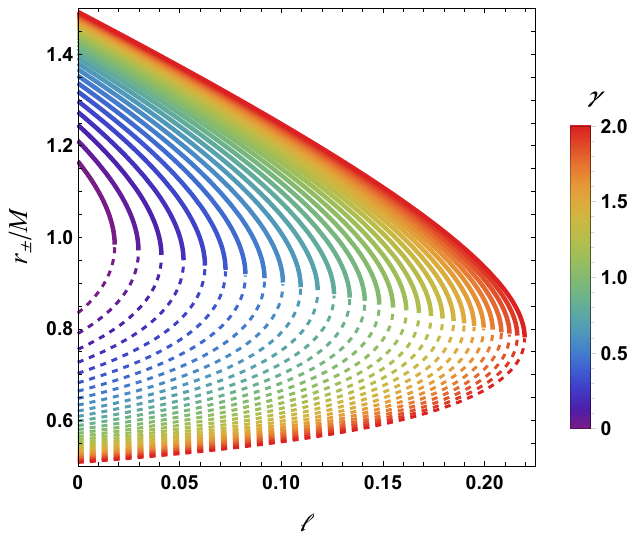}~~~~\\
\includegraphics[scale=0.78]{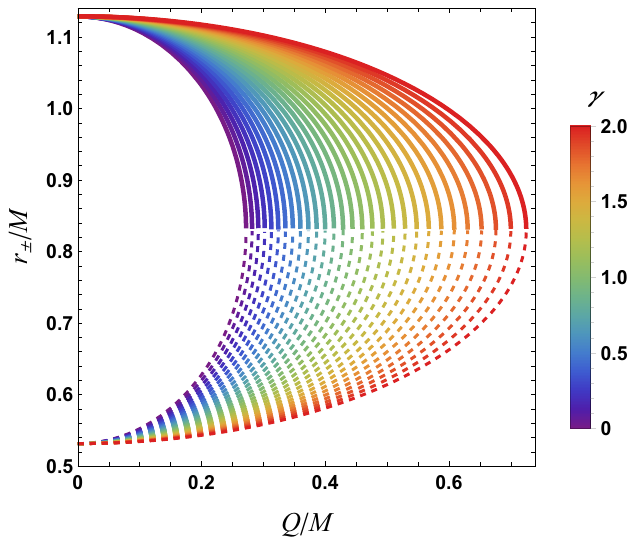}~~~
\caption{Event (solid curves) and Cauchy (dashed curves) horizons of the RKRBHs for various values of the tracless field $\gamma$, against $a/M$, $\ell$ and $Q/M$.} \label{extt}
\end{figure}

To shed light on the relevant horizon structure for the Newmann-Kerr-like BH solution, Figure~\ref{Figps} illustrates the associated parameter space $m_1=(\gamma, a/M)$, $m_2=(\gamma,\ell)$ and $m_3=(\gamma, Q/M)$. Effectively, the two-dimensional cross-sections reveal and highlight the extreme separation in each module space $m_i$ of the parameter model. Given that the sector of traceless/conformal matter managed by the ModMax model produces a modified energy-momentum tensor, the effective electromagnetic contribution appears multiplied by $e^{-\gamma}$, and the Kalb-Ramond sector induces denominators dependent on $\ell$ in the metric coefficients; these two effects reshape the extremal curve in a non-trivial manner. In concrete terms, this means that the permissible observable values of spin $a$ are limited for large values of $\ell$ or $Q$, while increasing $\gamma$ opens up certain parts of the parameter space by filtering out charge effects. The panel therefore provides the natural theoretical prior used in shadow- and horizon-based parameter estimation. The coloured regions represent black holes with two horizons, while the black curves mark extremal cases where $r_+ = r_-$. The white regions correspond to spacetimes without horizons (naked singularities). The RKRBH metric~\eqref{xian} describes a non-extremal BH for $r_+ > r_-$ and an extremal BH with $r_+ = r_-$. Graphically, solid (dashed) curves show $r_{+}$ ($r_{-}$) and illustrate how the separation $r_{+}-r_{-}$ closes near extremal parameter combinations (see Figure \ref{extt}). The figures demonstrate the analytic expectation that the electrostatic term appears as $e^{-\gamma}Q^{2}/(1-\ell)^{2}$ in the metric function $\Delta_r(r)$, so that increasing $\gamma$ reduces the magnitude of the charge term and tends to increase $r_{+}$ (for fixed $Q$), whereas increasing $\ell$ enhances the relative importance of the charge term through the $(1-\ell)^{-2}$ factor and thus reduces $r_{+}$. The presence of traceless (conformal) matter in the action modifies the radial stress components and thus slightly alters the behaviour of the inner horizon $r_{-}$, producing subtle non-monotonicities visible for intermediate values of the parameters.

\begin{figure*}[hbt!]
\centering
\begin{tabular}{c c c c}
\includegraphics[scale=0.48]{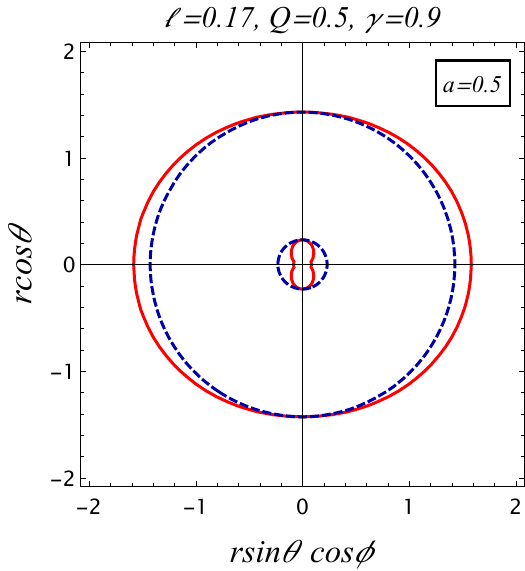}&
\includegraphics[scale=0.48]{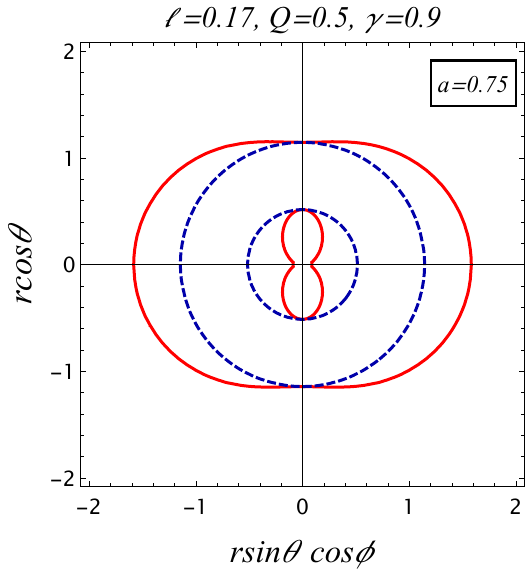}&
\includegraphics[scale=0.48]{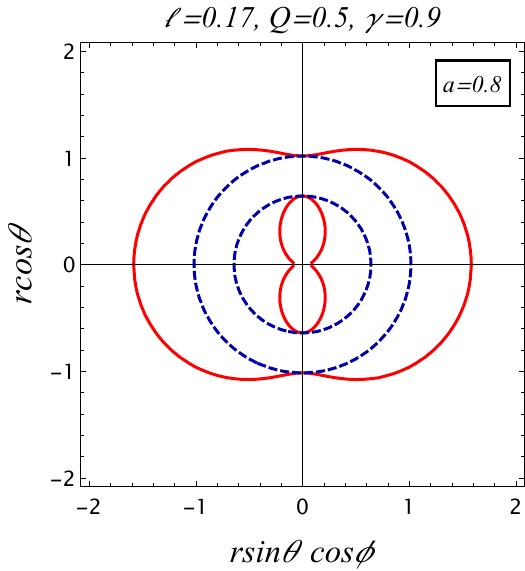}&
\includegraphics[scale=0.48]{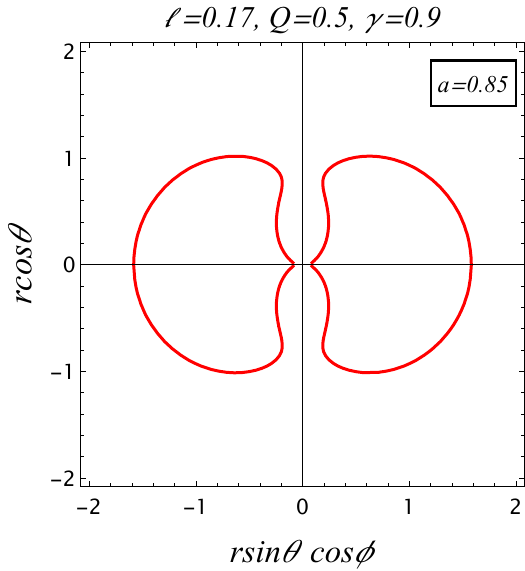}\\
\includegraphics[scale=0.48]{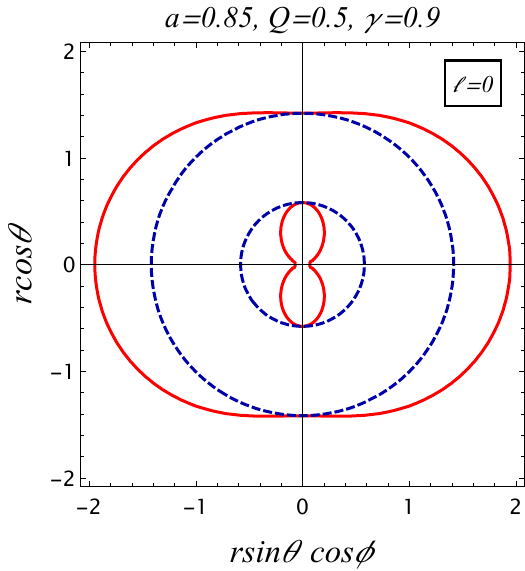}&
\includegraphics[scale=0.48]{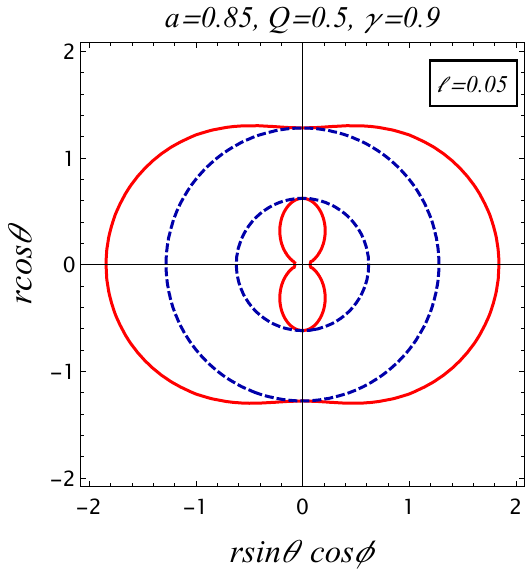}&
\includegraphics[scale=0.48]{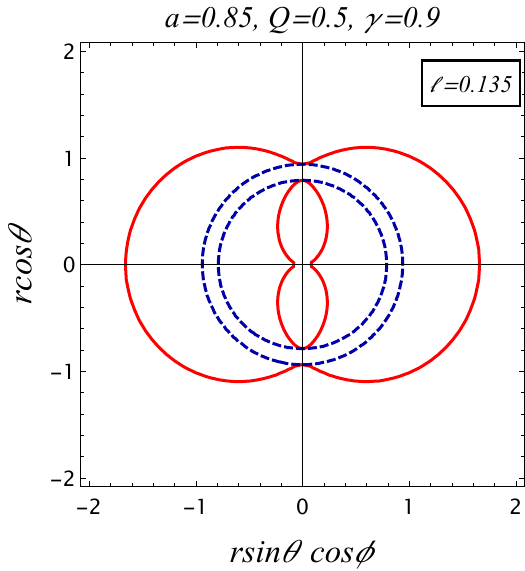}&
\includegraphics[scale=0.48]{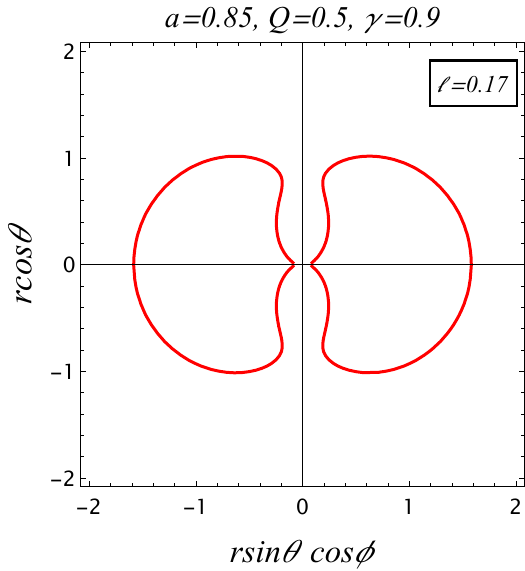}\\
\includegraphics[scale=0.48]{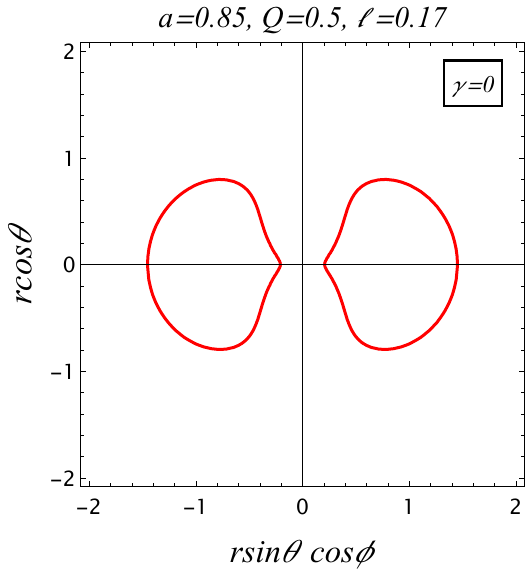}&
\includegraphics[scale=0.48]{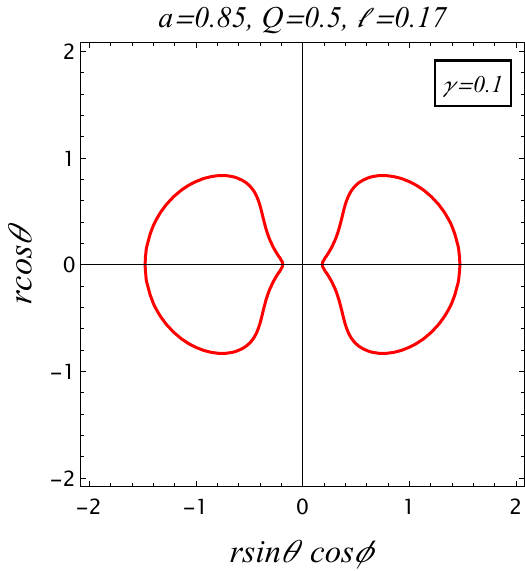}&
\includegraphics[scale=0.48]{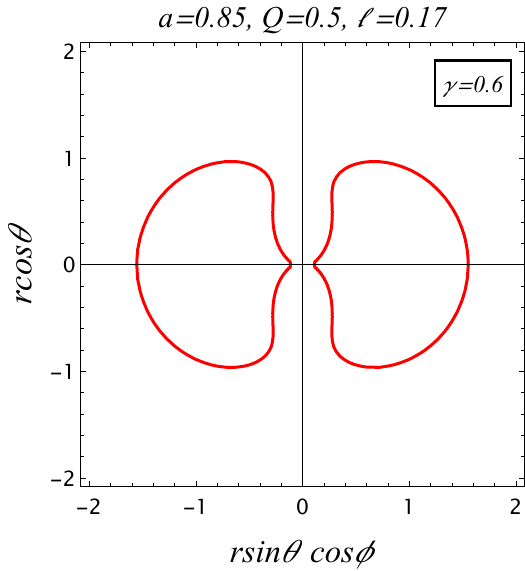}&
\includegraphics[scale=0.48]{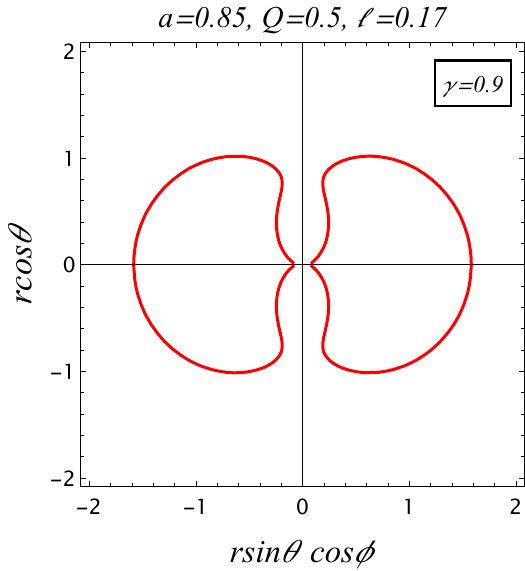}\\
\includegraphics[scale=0.48]{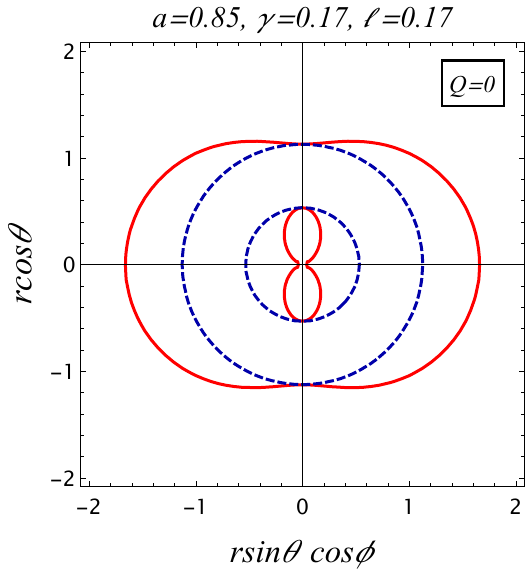}&
\includegraphics[scale=0.48]{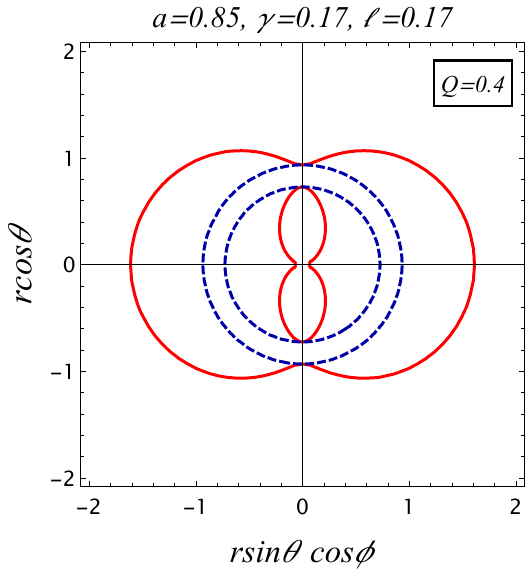}&
\includegraphics[scale=0.48]{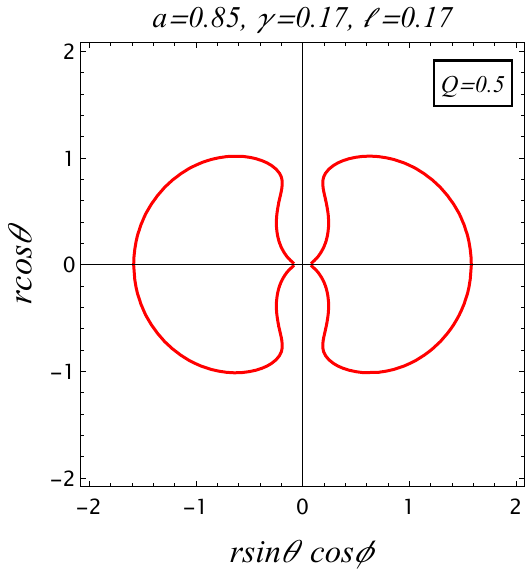}&
\includegraphics[scale=0.48]{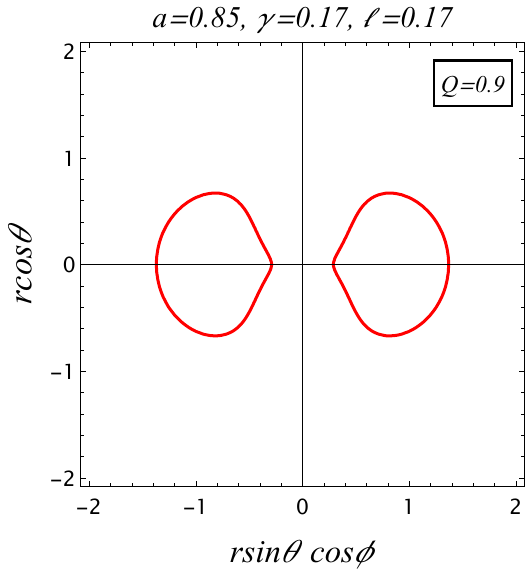}
\end{tabular}
\caption{The cross-section of the event horizon (outer blue curve), SLS (outer red dotted curve) and ergoregion of the RKRBHs for different values of  parameters $Q$, $\ell$, $\gamma$ and $a$ respectively. The increased values of  parameters $a$ (top), $\ell$ (second row), $\gamma$ (third row), and $Q$ (bottom) lead to disconnected event horizons. }\label{Figergo}
\end{figure*}

The static limit surface (SLS) and the ergosphere of the RKRBH are defined by $g_{tt} = 0$, which admits two roots, viz $r_{\pm}^{SLS}$. The region $r_+^{EH} < r < r_+^{SLS}$ describes the \textit{ergosphere}, where the Killing field $\xi^a = (\partial/\partial t)^a$ turns out to be spacelike, forcing observers to co-rotate. Its shape is sensitive to the spin $a$ and parameters $\ell$, $\gamma$, and $Q$. Considering Cartesian coordinates $x=r\sin\theta\cos\phi$,$y=r\sin\theta\sin\phi$, $z=r\cos\theta$ and the axisymmetry of the rotating BHs, in Figure~\ref{Figergo}, the cross-sections show the event horizon and the SLS and quantify the ergoregion's deformation with $a,\ell,\gamma,Q$. Spin $a$ continues to be the primary source of ergoregion asymmetry; however, the KR parameter $\ell$ systematically alters the SLS radius through the factor $(1-\ell)^{-1}$. This can lead to the formation of disconnected or highly deformed ergoregions when extreme values are considered. The ModMax parameter $\gamma$ influences the contribution of electromagnetic stresses to frame dragging through the effective rescaled charge. Additionally, the traceless-field coupling ensures that the electromagnetic/KR sector introduces a conformal-like, anisotropic stress. This, in turn, results in enhanced angular dependence of the SLS compared to the Kerr-Newman solution, which is directly observable in the panels. The panel therefore highlights the astrophysical implications relevant to energy extraction processes and GRMHD initial conditions.

%Henceforth, both models will be treated on equal footing and collectively referred to as RQBH.    
%{\color{green}\bf Discussion regarding static limit surface cross-section.}

\section{Null geodesic around the RKRBHs and shadows}\label{sec:geodesics}
In this section we investigate the null geodesics of the rotating black hole solution and study the impact of the Lorentz-violating parameter $\ell$ and of the traceless-conformal sector on the angular size of the shadow. Since the shadow boundary is determined by photon orbits, we begin by analysing the null geodesic equations in the background metric $g_{\mu\nu}$.

The geometry defined by Eq.~\eqref{xian} possesses two Killing symmetries, generated by the timelike Killing field $\xi_{(t)}^a=(\partial/\partial t)^a$ and the axial Killing field $\xi_{(\phi)}^a=(\partial/\partial\phi)^a$. Consequently, null geodesics admit two conserved quantities associated with these isometries, namely the conserved energy $E=-\xi_{(t)}^a p_a$ and the axial angular momentum $L=\xi_{(\phi)}^a p_a$, where $p_a$ is the photon four-momentum~\cite{Wald:1984bk}. So one has
\begin{equation}
\begin{split}
&E:=-g_{t t} \frac{\mathrm{d} t}{\mathrm{~d} \tau }-g_{t\phi}\frac{\mathrm{d} \phi}{\mathrm{~d} \tau },\\
&L:=g_{\phi \phi} \frac{\mathrm{d} \phi}{\mathrm{d} \tau }+g_{t\phi}\frac{\mathrm{d} t}{\mathrm{~d} \tau },\label{EL}
\end{split}
\end{equation}
where $\tau$ represents the affine parameter.  Geodesics and orbits in Kerr spacetime were initially examined by Carter \cite{Carter:1968rr} through the Hamilton-Jacobi equation. This analysis resulted in the identification of an additional constant of motion: the Carter constant, which arises from the separability of the equation. The constant can also be derived from the Killing tensor $K_{ab}$ \cite{Wald:1984bk}. Refs. \cite{Azreg-Ainou:2014pra, Shaikh:2019fpu, Junior:2020lya} demonstrate that stationary spacetimes generated via the NJA, when expressed in Boyer-Lindquist-like coordinates, also permit Hamilton-Jacobi separability. The Hamilton-Jacobi equation takes the form
\begin{equation}
\frac{\partial\mathcal{S}}{\partial\tau}+\frac{1}{2} g^{\mu\nu} p_\mu p_\nu=0, \quad p_\mu=\frac{\partial\mathcal{S}}{\partial x^{\mu}},
\label{eq:HJ_0}
\end{equation}
where  $\mathcal{S}$ is the Jacobi action, and $\bm{p}$ is the conjugate momentum covector. The Jacobi action can be expressed in a separable form as
\begin{equation}
S=\frac{1}{2} m^{2} \tau -E t+L \phi+S_{r}(r)+S_{\theta}(\theta),  \label{s32}            
\end{equation}
where $m$ represents the mass of the test particle, and it takes the value of 0 for the photon. By considering this ansatz, together with the line element defined in \eqref{xian} and the definitions expressed by \eqref{eq:HJ_0}, the Hamilton-Jacobi equation can be expressed in the following form:
\begin{align}
&-\frac{1}{\rho^2\Delta_r}\left[E\left(r^2+a^2\right)-aL\right]^2+\frac{p_\theta^2}{\rho^2}+\frac{1}{\rho^2\sin^2\theta}\left(aE\sin^2\theta-L\right)^2 +\frac{\Delta_r\, p_r^2}{\rho^2}=-m^2.
\label{eq:HJ_4}
\end{align}

For null geodesics, which describe the trajectories of massless particles such as photons, it is standard practice to set the rest mass $m = 0$. This reflects a fundamental characteristic of null geodesics in GR, where the condition $\mathrm{d}s^2 = 0$ holds true. We can deduce the equation
\begin{align}
\mathcal{K}\equiv &\Delta_r\, p_r^2 - \frac{1}{\Delta_r}\left[E\left(r^2 + a^2\right) - aL\right]^2 = -\left[p_\theta^2 + \frac{1}{\sin^2\theta}\left(aE\sin^2\theta - L\right)^2\right],
\label{eq:HJ-5}
\end{align}
where we have separated the \( r \)-dependent segment of the equation from the \( \theta \)-dependent part, using a constant \( \mathcal{K} \) defined as
\begin{equation}
\mathcal{K}= \mathbb{Q}+\left(aE - L\right)^2,
\label{eq:mK}
\end{equation}
with $\mathbb{Q}$  representing Carter's constant \cite{Carter:1968rr}.

Based on this method, the equations for \( p_r \) and \( p_\theta \) are obtained as
\begin{eqnarray}
p_r^2 &=& \frac{1}{\Delta_r^2} \left[E\left(r^2 + a^2\right) - a L\right]^2 - \frac{1}{\Delta_r}\left[Q + \left(aE - L\right)^2\right], \label{eq:pr} \\
p_\theta^2 &=& Q + a^2 E^2 \cos^2\theta - L^2 \cot^2\theta. \label{eq:ptheta}
\end{eqnarray}
As evident from the foregoing, the Hamilton-Jacobi equation for null geodesics is fully separable in the rotating line element \eqref{xian}. We note that the standard NJA generally guarantees separability only when the resulting rotating metric can be expressed in Boyer-Lindquist coordinates. Thus, substituting Eq.~\eqref{s32} into Eq.~\eqref{eq:HJ_0} and separating the radial and polar dependences yields the usual Carter separation constant, from which one obtains the decoupled null geodesic equations for $r$ and $\theta$ determined by the background metric \eqref{eq:ds_kerr_like} as
\begin{align}
\left(\rho^2\frac{d r}{d \tau }\right)^{2}&=\left[\left(r^2+a^{2}\right)E-a L\right]^{2}-\Delta_r\left[C+(L-a  E)^{2}\right] \equiv R(r), \label{likenull3}\\
\left(\rho^2 \frac{d \theta}{d \tau }\right)^{2}&=C+a^{2} E^{2} \cos^{2} \theta-L^{2}\cot^{2}\theta \equiv\Theta(\theta), \label{likenull4}
\end{align}
where $C$ is the Carter constant, and we have used $\frac{\partial S}{\partial x^{\mu}} = p_{\mu}$ and $p^{\mu} = g^{\mu \nu} p_{\nu} = \dot{x}^{\mu} = \frac{d x^{\mu}}{d \tau}$ with \(p^{\mu}\) being the  momentum. Solving Eq.~\eqref{EL} with the metric \eqref{xian} yields
\begin{align}
\rho^2 \frac{d t}{d \tau} &=  a \Big(L - a E\,\sin^2\theta\Big) \Bigg. 
\Bigg. + \frac{r^2 + a^2}{\Delta_r(r)} \Big[\Big(r^2 + a^2\Big) E - a L\Big] ,\\
\rho^2 \frac{d \phi}{d \tau }&= {\left ( \frac{L}{\sin ^{2} \theta}-a E\right.} \left.+ \frac{a}{\Delta_r}\left[\left(r^2+a^{2}\right) E-a L\right]  \right ) .\label{likenull2}
\end{align}
The null geodesic equations \eqref{likenull3}-\eqref{likenull2} describe how the photons move around the rotating BHs. It is convenient to introduce two impact parameters \cite{Bardeen:1973}
\begin{equation}
\xi=\frac{L}{E}, \quad \eta=\frac{C}{E^{2}},\label{eq:twoimpactpara}
\end{equation}
which are the gauge-invariant constants of motion, and we have also defined
\begin{subequations}
\begin{eqnarray}
R(r) &=& \left(r^2 + a^2 - a\xi\right)^2 - \Delta_r \left[\eta + \left(a - \xi\right)^2\right], \label{eq:mR} \\
\Theta(\theta) &=& \eta + \cos^2\theta \left(a^2 - \frac{\xi^2}{\sin^2\theta}\right). \label{eq:Theta}
\end{eqnarray}
\end{subequations}
It is essential to rely on the conditions \( R(r) \geq 0 \) and \( \Theta(\theta) \geq 0 \) for the null trajectories to exist in the exterior geometry of the BH.
\begin{figure}[hbt!]
  	\centering
  	\includegraphics[width=7.5cm,height=7cm]{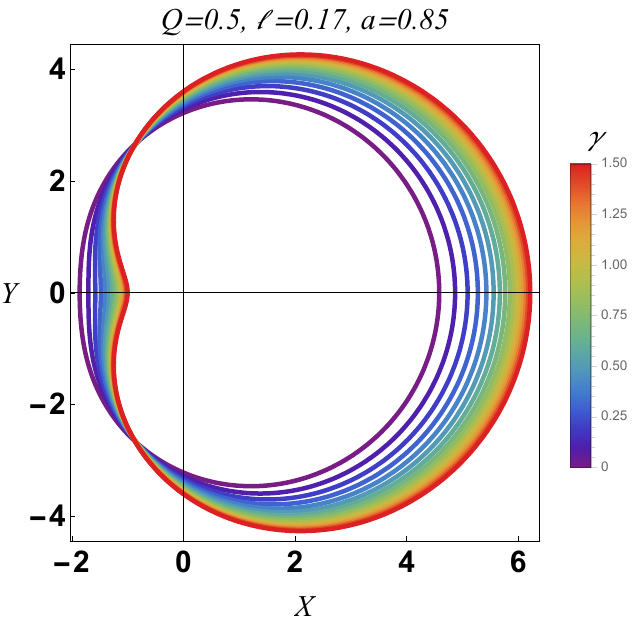}~~~
  	\includegraphics[width=7.5cm,height=7cm]{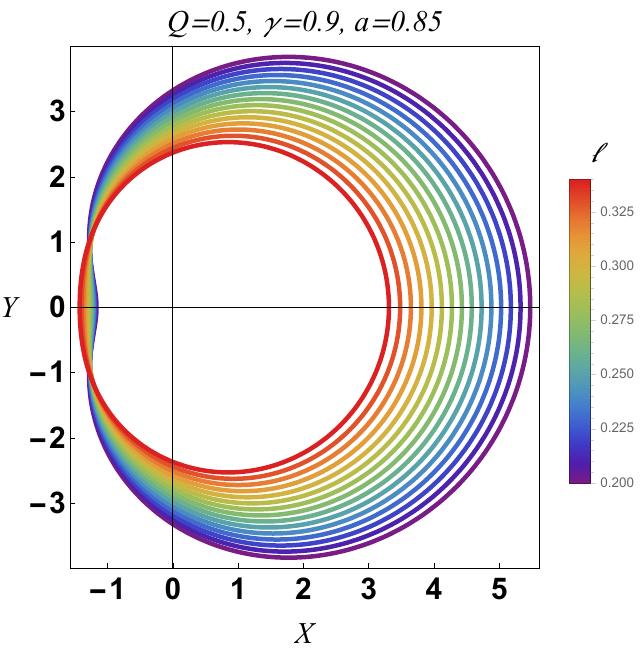}~~~~\\
\includegraphics[width=7.8cm,height=7cm]{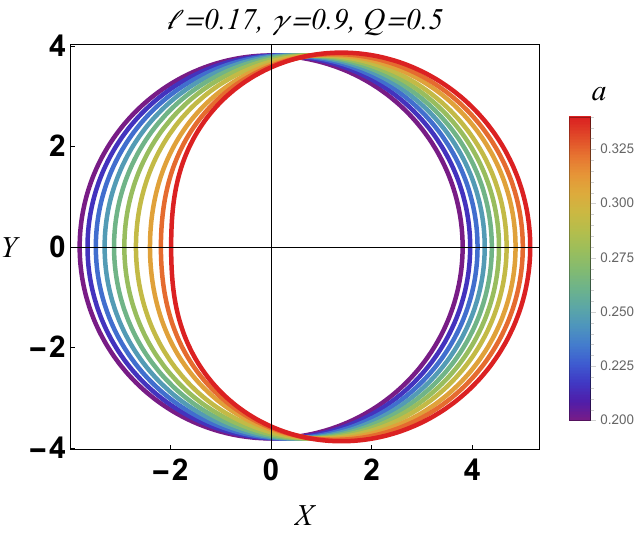}~~~
\includegraphics[width=7.5cm,height=7cm]{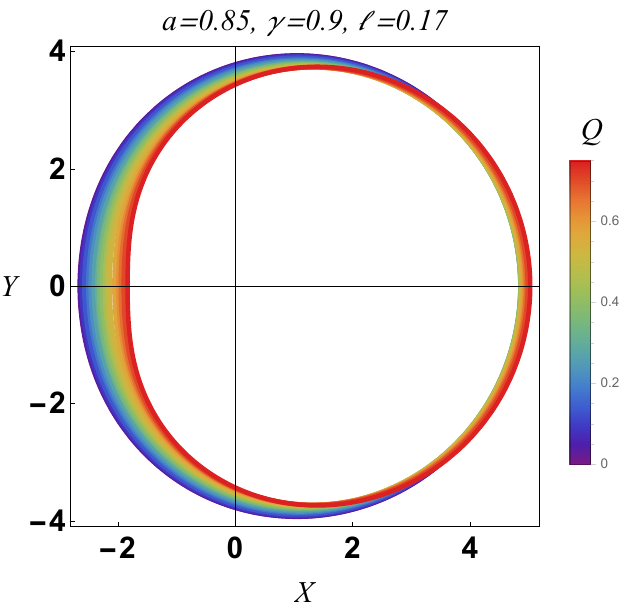}
  	\caption{Shadow silhouettes of the RKRBHs with varying $\gamma$ (top left), $\ell$ (top right) $a$ (bottom left) and $Q$ (bottom right)   for an inclination angle $\theta = \pi/2$.}
  	\label{Figshh}
\end{figure}

\begin{figure}[hbt!]
  	\centering
  	\includegraphics[width=7.5cm,height=7cm]{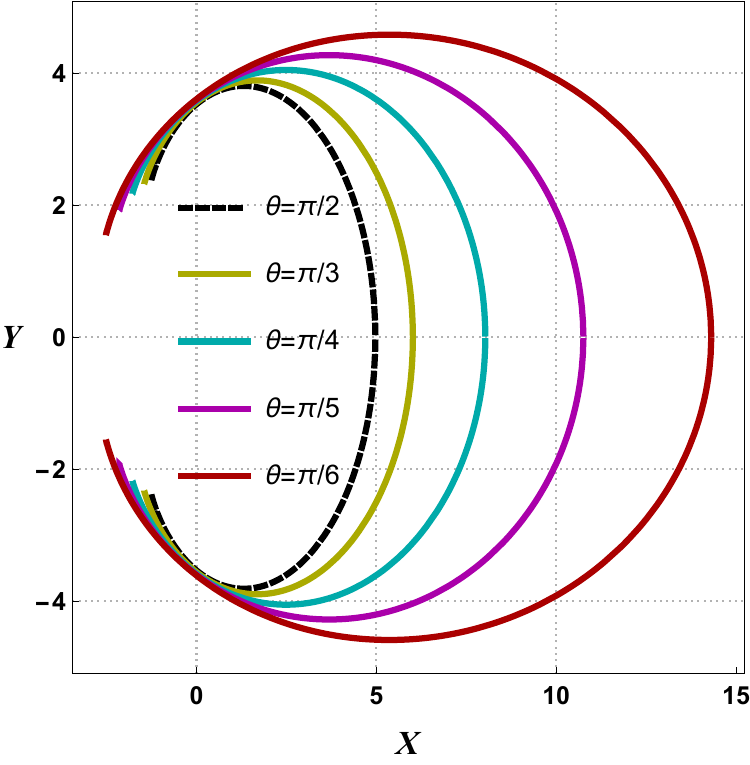}~~~
  	
  	\caption{Shadow silhouettes of the RKRBHs with varying the inclination angle $\theta$ for $M=1,\, Q=0.5,\,\gamma=0.1$ and $\ell=0.17$.}
  	\label{Figshh2}
\end{figure}

\subsection{Shadows of RKRBHs}
Following the approach in \cite{Bardeen:1973}, we consider a uniformly bright source positioned behind the BH, with an angular size significantly larger than that of the BH itself, and a distant observer located in front of it. Photons emitted by this source can follow one of three trajectories in the vicinity of a rotating BH, determined by their impact parameter: scattering orbits, capture orbits, or unstable orbits \cite{Bardeen:1973, Amir:2016cen}. In scattering orbits, photons are deflected and successfully escape to infinity. Conversely, in capture orbits, they are drawn into the BH. Unstable orbits, which exist between these two categories, are particularly sensitive to minor perturbations that can influence whether a photon ultimately escapes or is captured. Physically, the set of unstable photon orbits constitutes a critical surface whose projection onto the observer's sky marks the transition between photon capture and escape, thereby manifesting as the boundary curve of the BH shadow.
The condition for unstable photon orbits is given in \cite{Hioki:2009na}
\begin{equation}
R\left(r_{s}\right) =
\left.\frac{d R(r)}{d r}\right|_{r=r_{s}}=0 ,  \quad \left.\frac{d^{2} R(r)}{d r^{2}}\right|_{r=r_{s}} \geq0 ,
\label{R}
\end{equation}
where $r_s$ represents the radius of the photon sphere.  By expressing $R$ from Eq.~\eqref{eq:mR} in terms of the impact parameters $\xi$ and $\eta$ as given in Eq.\eqref{eq:twoimpactpara}, and solving the first part of Eq.~\eqref{R} for $\xi$ and $\eta$ as functions of $r_s$, we obtain the desired relations.

In fact, spherical photon orbits located at a fixed radius $r_p$ satisfy the conditions $R(r_p) = 0 = R'(r_p)$. Using the structure of $R$ provided in Eq.~\eqref{eq:mR}, these conditions yield a set of solutions for the impact parameters.

\begin{eqnarray}
&& \xi = \left(\frac{r^2+a^2}{a}\right)_{r_s}, \quad \eta = -\left(\frac{r^4}{a^2}\right)_{r_s},
\label{eq:xip_1}
\end{eqnarray}
and 
\begin{eqnarray}
&& \xi = \left(\frac{r^2+a^2}{a}-\frac{4r\, \Delta_r}{a \Delta_r'}\right)_{r_p}, \quad \eta = \left(\frac{16\Delta_r\, r^2}{(\Delta_r')^2}-\frac{1}{a^2}\left[r^2-\frac{4r\, \Delta_r}{\Delta_r'}\right]^2\right)_{r_p}.\label{eq:etap_2}
\end{eqnarray}

These expressions determine the critical impact parameter for photons on constant-$r$ trajectories. However, as seen from Eq.~\eqref{eq:xip_1}, the condition $\eta < 0$ yields an unphysical outcome. Therefore, we discard the solutions from Eq.~\eqref{eq:xip_1} and instead adopt those from Eq.\eqref{eq:etap_2} as the critical conserved quantities for spherical photon orbits. In particular, for equatorial orbits where $\theta = \pi/2$, Eqs.\eqref{likenull4} and \eqref{eq:Theta} imply that $\eta = 0$.

Photons with the critical impact parameter are confined to unstable, circular null orbits that collectively produce the lensed ``photon ring''. In the Kerr geometry these critical impact parameters depend on the photon direction with respect to the spin axis, and as a consequence the apparent shadow boundary is generically noncircular~\citep{Kerr:1963ud,Bardeen:1973tla}.  Although the resulting azimuthal distortion of the shadow is typically small (of the order of a few percent, often quoted as $\lesssim 4\%$), it can be measurable under favourable observational conditions~\citep{Takahashi:2004xh,Johannsen:2010ru}. These unstable photon trajectories, which separate geodesics that are captured by the hole from those that escape to infinity, have been analysed for both black holes and naked singularities~\citep{Wilkins:1972rs,Johnston:1974pn,Teo:2020sey}.  In axisymmetric spacetimes the equatorial plane generally admits two circular photon orbits: a prograde orbit (co-rotating with the hole) at the smaller radius \(r_{p}^{-}\), and a retrograde orbit (counter-rotating) at the larger radius $r_p^+$.  Frame dragging (the Lense-Thirring effect) forces prograde photons onto tighter orbits, thereby increasing their angular momentum, whereas retrograde photons must reside at larger radii to balance the loss of angular momentum~\citep {Bardeen:1972fi,Bardeen:1975zz,Teo:2020sey}. Motion transverse to the equatorial plane becomes possible whenever the separation constant $\eta_{c}$ is positive; for $\eta_{c}=0$, the photon motion is confined to planar (equatorial) trajectories. Consequently, the radii $r_{p}^{\pm}$ of the prograde and retrograde circular photon orbits are determined by the roots of the condition $\eta_{c}=0$ evaluated in the equatorial plane.
In particular, the radii for prograde and retrograde circular photon orbits in the equatorial plane, for the Kerr case, are given by \citep{Teo:2020sey}
\begin{equation}\label{photonRKerr}
\begin{aligned}
r_p^- &= 2M\left[1 + \cos\left(\frac{2}{3} \cos^{-1}\left(-\frac{|a|}{M}\right)\right)\right], \\
r_p^+ &= 2M\left[1 + \cos\left(\frac{2}{3} \cos^{-1}\left(\frac{|a|}{M}\right)\right)\right],
\end{aligned}
\end{equation}
with $r_p^-$ and $r_p^+$ confined to the ranges $M \leq r_p^- \leq 3M$ and $3M \leq r_p^+ \leq 4M$, respectively. For a Schwarzschild black hole ($a = 0$), these two orbits merge into a single photon sphere of radius $r_p = 3M$. The inequality $r_p^- \leq r_p^+$ arises due to the Lense-Thirring effect \citep{Johannsen:2010ru}, while in the extremal Kerr case ($a = M$), the prograde orbit radius coincides with the event horizon, yielding $r_p^- = r_E = M$.

Next, we derive the relation between two constants, $\xi_c$ and $\eta_c$, and the observer's image plane coordinates, $X$ and $Y$. For an observer at the position ($r_*,\theta_*$). Consider an observer situated at infinity, with a sight angle $\theta$ between the line of sight and the BH's rotation axis. The contour of the BH shadow as seen by this observer can be described by the celestial coordinates, given by

\begin{align} X &= \lim {r{\ast} \rightarrow \infty}\left(-r_{\ast}^{2} \sin \theta_{\ast} \frac{d \phi}{d r}\bigg|{\theta \rightarrow \theta{\ast}}\right) = -\xi \csc \theta_{\ast}, \nonumber \\ \ Y &= \lim {r{\ast} \rightarrow \infty}\left(r_{\ast}^{2} \frac{d \theta}{d r}\bigg|{\theta \rightarrow \theta{\ast}}\right) = \pm \sqrt{\eta + a^{2} \cos ^{2} \theta_{\ast} - \xi^{2} \cot ^{2} \theta_{\ast}}, \label{eq:Y} \end{align}

In the second step of Eq. \eqref{eq:Y}, the null geodesic equations \eqref{likenull2} and \eqref{likenull3} were applied, while Eqs. \eqref{likenull4} and \eqref{likenull3} were used in deriving \eqref{eq:Y}. Notably, the undetermined factor $\Psi$, which frequently appears in the null geodesic equations, cancels out in the expressions for $X$ and $Y$. As a result, the rotating BH shadows are independent of the choice of $\Psi$, as highlighted by \cite{Junior:2020lya}. If the observer is positioned on the equatorial plane of the rotating BHs with $\theta_{\ast} = \frac{\pi}{2}$, we obtain
\begin{equation}
\begin{split}
X =-\xi , \quad
Y = \pm \sqrt{\eta} .\label{XY}
\end{split}
\end{equation}
In what follows, we will focus on this case. It is not difficult to see that the obtained celestial coordinates are constrained by \(a\), \(Q\), $\gamma$ and \(\ell\). Therefore, in order to get accurate BH shadows, one needs first to constrain the parameters \(a\), \(Q\), $\gamma$ and \(\ell\). The same parameter space of $(Q,\gamma,a,\ell)$ allowing the existence of shadows cast by RKRBHs is plotted in Fig. \ref{Figshh}. The shadow silhouettes of the RKRBHs with varying the inclination angle $\theta$ for $M=1,\, Q=0.5,\,\gamma=0.1$ and $\ell=0.17$ is depicted in Fig.~\ref{Figshh2}.

Graphically speaking, the four subpanels isolate how each parameter deforms the photon capture region projected on the observer's sky. 
Increasing $a$ produces the familiar D-shaped displacement and asymmetry; nonzero $\ell$ introduces additional non circular distortions and small-scale kinks because the KR field modifies the effective geometry experienced by null geodesics (through $(1-\ell)$-dependent metric coefficients). 
The ModMax parameter $\gamma$, which enters the electromagnetic sector as an $e^{-\gamma}$ weighting of $Q$, acts to \emph{reduce} charge-induced shadow shrinkage for larger $\gamma$, i.e.\ it partially restores circularity for fixed $Q$. 
Given the traceless/conformal coupling of matter in the model, these effects are amplified in angular sectors where the traceless stress-energy departs most from isotropy; thus $\gamma$ and $\ell$ produce qualitatively distinct signatures that can when combined with a proper emission model be sought in EHT--like data.

The sequence of inclination angles demonstrates the well-known projection degeneracy: as the observer moves away from the equatorial plane the apparent asymmetry decreases. 
Importantly, the figure shows that the degeneracy between inclination $\theta$ and intrinsic deformations induced by $\ell$ and $Q$ is nontrivial because the traceless-field contributions break spherical symmetry at the stress--energy level; hence an object with modest intrinsic deformation at high inclination can mimic a more strongly deformed but more face on configuration. 
The ModMax screening parameter $\gamma$ again moderates charge-driven deviations, therefore accentuating parameter degeneracies $(a,\ell,Q,\gamma,\theta)$ that must be explored jointly in any observational inference.

\section{Shadow Observables and Parameter Estimation}\label{sec:Shadow}
Estimating black hole parameters from observed shadows is a cornerstone of contemporary black hole astrophysics.  We adopt the shadow-analysis framework introduced by Hioki and Maeda \cite{Hioki:2009na}, which has been used with EHT data to place tight constraints on spin and inclination \cite{Kumar:2018ple} and later extended to hairy Kerr spacetimes and other non-Kerr families \cite{Afrin:2021imp,Afrin:2023uzo,Afrin:2021ggx,Afrin:2024khy,Ali:2024ssf}.  Although the EHT images of M87* and Sgr A* are broadly consistent with Kerr expectations \citep{EventHorizonTelescope:2019dse,EventHorizonTelescope:2022wkp}, modified-gravity black holes (e.g. RKRBHs) remain largely untested observationally, motivating the present comparison.  For direct comparison with the EHT constraints, we assume an equatorial observer, \(\theta_0=\pi/2\), as well as some other inclination angle.

\paragraph{Hioki--Maeda observables.}
Following Hioki and Maeda \cite{Hioki:2009na} we characterize the shadow by two geometrical observables: the reference radius \(R_s\) (shadow size) and the distortion parameter \(\delta\) (shadow asymmetry).  Let the three intersection points between the shadow and the reference circle be the top \((X_t,Y_t)\), bottom \((X_b,Y_b)\) and rightmost \((X_r,0)\) points, and denote the leftmost points of the shadow and of the reference circle by \((X_l,0)\) and \((X_l',0)\), respectively \citep{Ghosh:2020ece}.  The circle radius \(R_s\) that passes through \((X_t,Y_t)\), \((X_b,Y_b)\) and \((X_r,0)\) is given by
\begin{equation}\label{Rs_HiokiMaeda}
R_s \;=\; \frac{(X_t-X_r)^2+Y_t^2}{2\,\big|X_t-X_r\big|}\,,
\end{equation}
and the dimensionless distortion parameter is defined as
\begin{equation}\label{deltas_HiokiMaeda}
\delta \;=\; \frac{X_l' - X_l}{R_s}\,,\qquad
X_l' \;=\; X_r - 2R_s,
\end{equation}
so that \(\delta\) measures the fractional displacement of the shadow's leftmost edge relative to the reference circle.  We compute the coordinates \((X_i,Y_i)\) from the photon-impact-parameter map for RKRBH spacetimes (evaluated at \(\theta_0=\pi/2\)) and use Eqs.~\eqref{Rs_HiokiMaeda}--\eqref{deltas_HiokiMaeda} to extract spin/inclination and to place observational bounds on the model parameters.

Thus, $R_s$ estimates the shadow's overall size, while $\delta$ captures its asymmetry. These observables are formally defined as~\citep{Hioki:2009na}:
\begin{eqnarray}
R_s &=& \frac{(X_t - X_r)^2 + Y_t^2}{2|X_r - X_t|}, \label{Rs}
\end{eqnarray}
where the relations $X_b = X_t$ and $Y_b = -Y_t$ hold, and
\begin{eqnarray}
\delta &=& \frac{|X_l - X_l'|}{R_s}, \label{deltas}
\end{eqnarray}
with subscripts $r$, $l$, $t$, and $b$ representing the shadow boundary's right, left, top, and bottom edges.

\begin{figure*}[hbt!]
\begin{tabular}{c c}
\includegraphics[scale=0.57]{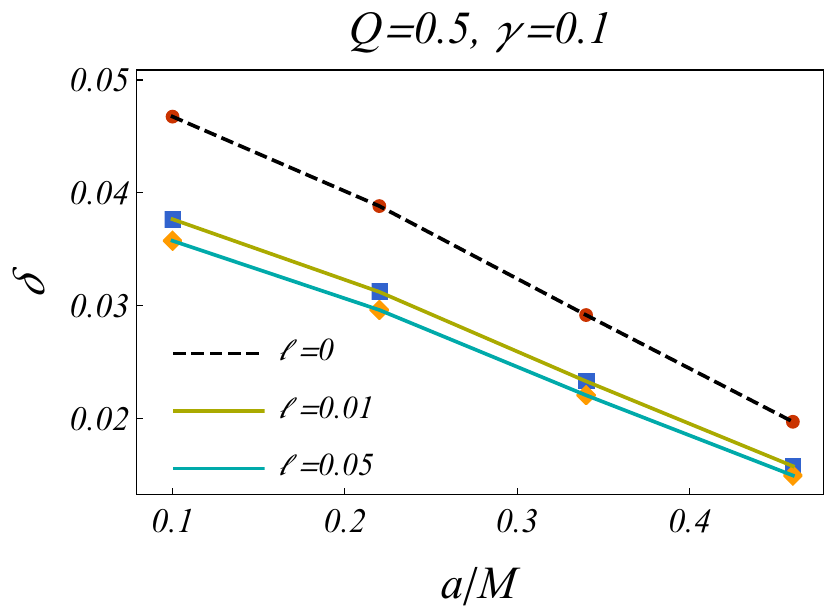}
\includegraphics[scale=0.57]{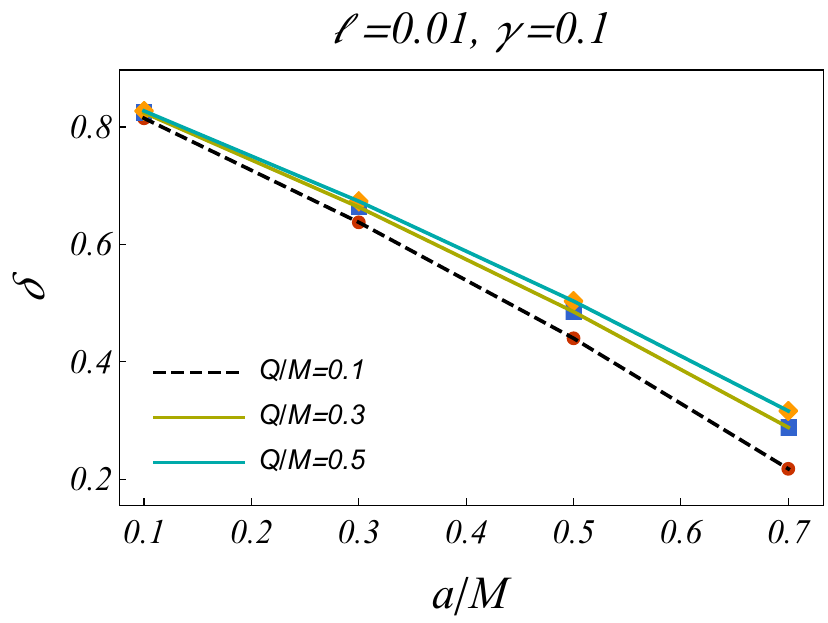}
\end{tabular}
\caption{Plots of the shadow observable $\delta$ versus $a/M$ in the RKRBH spacetime for various values of Lorentz-violating parameter $\ell$ and the electric charge $Q$ with $\theta=\pi/2$.  }
\label{Figparameterestimation1}
\end{figure*}

Tabs~\ref{parameter_table1}-\ref{parameter_table2} summarise the computed shadow diagnostics (for example, areal radius \(A\), characteristic radius \(R_s\), distortion parameter \(\delta\), oblateness \(D\), and centroid shift) evaluated on a dense grid in the intrinsic model parameters \((a/M,\;Q/M,\;\gamma,\;\ell)\). The numerical entries make two points particularly clear. First, the ModMax screening parameter \(\gamma\) enters primarily as a multiplicative attenuation of the electromagnetic imprint: Increasing \(\gamma\) systematically reduces the charge-driven reductions in both \(R_s\) and \(A\), and therefore moves the row/column values closer to the uncharged (Kerr-like) baseline. Second, the Kalb-Ramond amplitude \(\ell\) induces a distinct class of distortions that are not captured by a simple rescaling of the charge: For fixed \(\mathcal{Q}_{\rm eff}=e^{-\gamma}Q^2/(1-\ell)^2\), the \(\ell\)-dependence produces angularly asymmetric changes in \(\delta\) (see Fig.~\ref{Figparameterestimation1}) and the centroid shift because \(\ell\) alters metric prefactors and the angular structure of the lensing potential. Across the table one therefore observes that (i) \(Q\) primarily controls the overall shrinkage of the photon capture scale, (ii) \(\gamma\) controls how strongly that shrinkage manifests observationally (screening), and (iii) \(\ell\) controls qualitative shape changes and higher-order moments.

\begin{figure}[hbt!]
  	\centering    \includegraphics[scale=0.57]{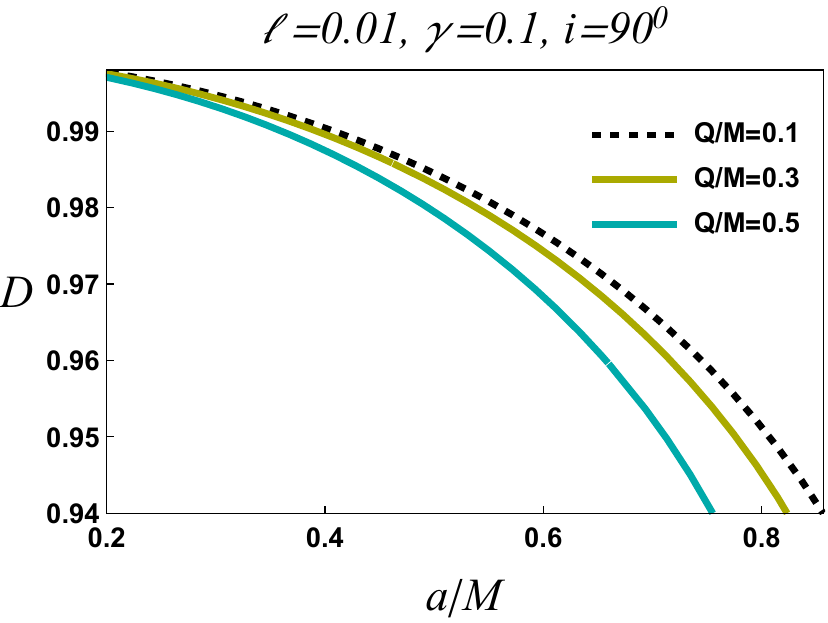}~~~
\includegraphics[scale=0.57]{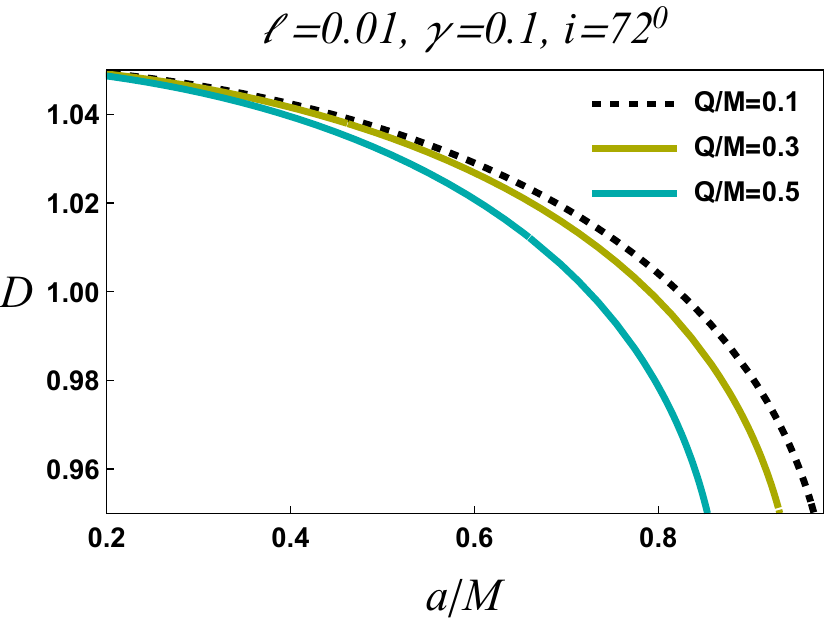}~~~~\\
\includegraphics[scale=0.57]{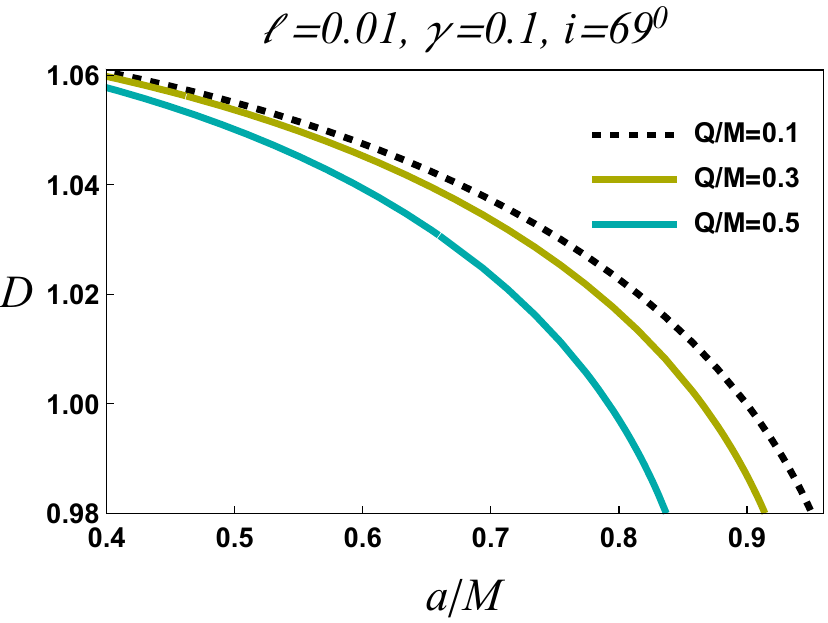}~~~
\caption{Variation of the shadow observable  $\mathfrak{D}$ against the spin $a/M$ for various values the charge $Q$ and inclination angle $\theta_0$.}\label{Figparameterestimation2}
\end{figure}

\begin{figure}[hbt!]
  	\centering
\includegraphics[scale=0.57]{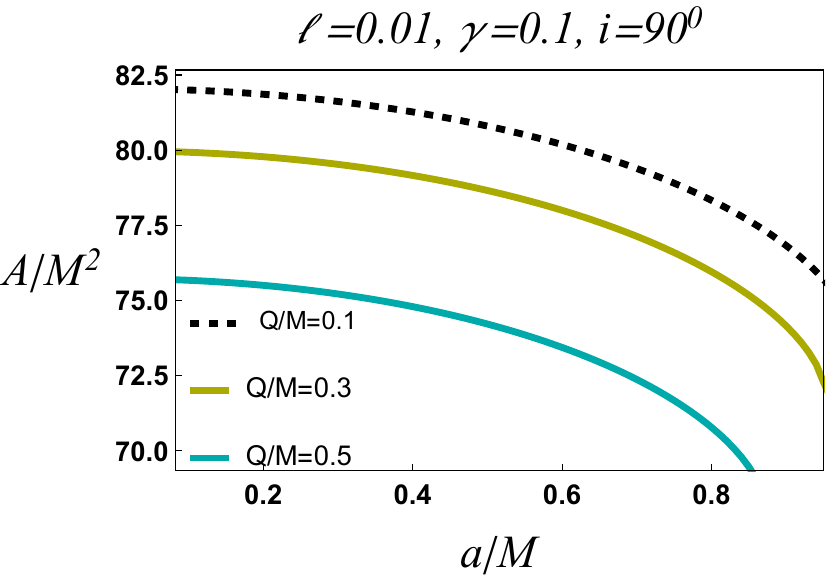}~~~
\includegraphics[scale=0.57]{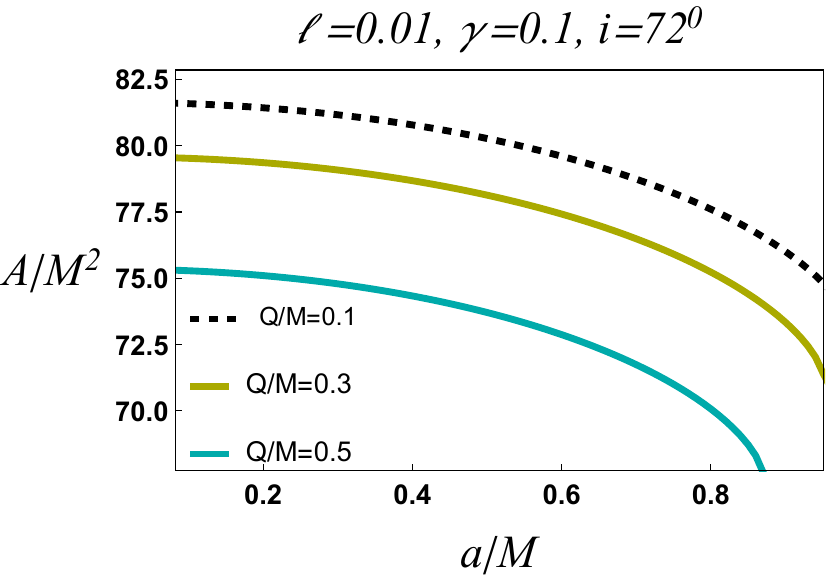}~~~~\\
\includegraphics[scale=0.57]{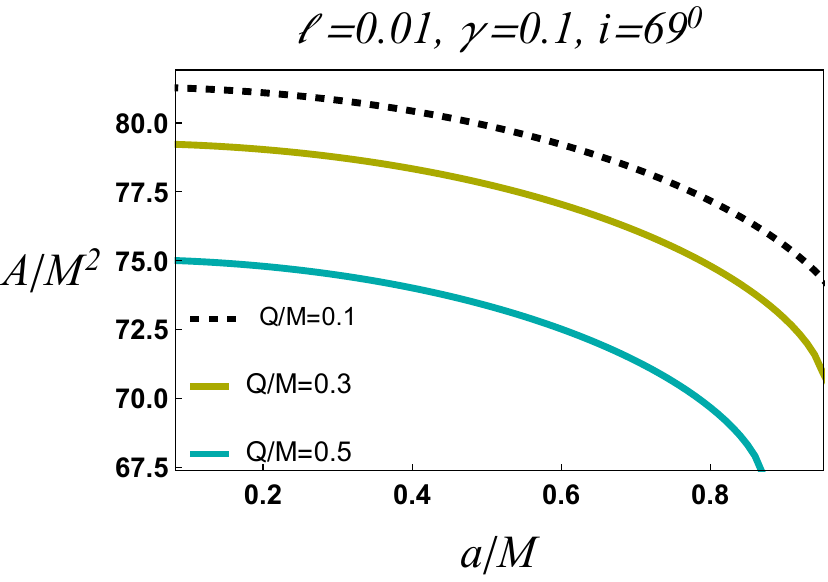}~~~
\caption{Variation of the shadow observable  $A/M^2$ against the spin $a/M$ for various values the charge $Q$ and inclination angle $\theta_0$.}\label{Figparameterestimation3} 
%The intersection points provide estimates of the parameters $a$ and $b_0$ when the values of $A_s$ and  $\mathfrak{D}_s$ for the RQBH }\label{Figparameterestimation3}
\end{figure}

\paragraph{Kumar and Ghosh method.}
Methods that build on Hioki--Maeda \citep{Hioki:2009na}  including analytic extensions \citep{Tsupko:2017rdo} and techniques to discriminate spinning MTG shadows from Kerr \citep{Tsukamoto:2014tja} rely on implicit shape symmetries that may break down for the irregular silhouettes frequently produced in modified-gravity scenarios \citep{Schee:2008kz,Johannsen:2015qca,Tsukamoto:2014tja,Abdujabbarov:2015xqa,Younsi:2016azx,Tsupko:2017rdo}.  Moreover, instrumental and reconstruction noise can perturb the apparent circularity of the image, biasing parameter inference \citep{Abdujabbarov:2015xqa,Kumar:2018ple}.  To mitigate these shortcomings, Kumar and Ghosh \citep{Kumar:2018ple} proposed a set of global, model-independent shadow observables well suited to irregular contours: the shadow area \(A\) and the oblateness \(\mathfrak{D}\).  The area is obtained from the parametric photon-impact-parameter map \((X(r_p),Y(r_p))\) as
\begin{equation}\label{Area}
A \;=\; 2\int Y(r_p)\,dX(r_p)
     \;=\; 2\int_{r_p^-}^{\,r_p^+} Y(r_p)\,\frac{dX(r_p)}{dr_p}\,dr_p,
\end{equation}
while the oblateness (measuring the horizontal-to-vertical extent) is defined by
\begin{equation}\label{Oblateness}
\mathfrak{D} \;=\; \frac{X_r - X_l}{Y_t - Y_b},
\end{equation}
where \(X_l,X_r,Y_t,Y_b\) denote the left, right, top and bottom shadow extrema, respectively.  For an equatorial observer these observables obey \(\sqrt{3}/2 \leq \mathfrak{D} < 1\), recovering \(\mathfrak{D}=1\) for Schwarzschild \((a=0)\) and \(\mathfrak{D}=\sqrt{3}/2\) for an extremal Kerr black hole \((a=M)\) \citep{Tsupko:2017rdo}.  In practice \(A\) is computed by numerically integrating the impact-parameter parametrisation and \(\mathfrak{D}\) by locating the shadow extrema; this pair of robust, global diagnostics reduces sensitivity to local irregularities and observational noise.  We apply these measures to RKRBH spacetimes (evaluated numerically at \(\theta_0=\pi/2\) and graphically at some angles $\theta_0$) and present variations of \(A\) and \(\mathfrak{D}\) parameter plane in Figs.~\ref{Figparameterestimation2}-~\ref{Figparameterestimation3}.

This panel quantifies the non-linear dependence of the shadow area on charge and spin. 
The displayed contours demonstrate that $Q$ produces a pronounced shrinking of $A$ via the metric term $e^{-\gamma}Q^{2}/(1-\ell)^{2}$, while spin primarily redistributes area into asymmetry rather than dramatically changing the total areal size for moderate $a$.  Notably, the presence of the traceless-field alters the curvature of constant-$A$ contours because it adds direction-dependent stress components that slightly bias the area away from the Kerr--Newman expectation; this is precisely the effect exploited in the parameter-estimation table.

\begin{table*}[hbt!]
\begin{tabular}{|p{1.8cm}|p{1.8cm}|p{1.8cm}|p{1.8cm}| }
\hline 
$R_s$ & $\delta$ & $a$/M &  $\ell$\\
\hline \hline
3.92587 &0.0380635 &0.88  &0  \\ 
\hline 
3.89322 &0.0401882 &0.89  &0.01  \\ 
\hline
3.8608& 0.0424567 & 0.9  & 0.02 \\
\hline
3.82864 & 0.044882 & 0.91 & 0.03 \\
\hline
3.79678& 0.0474785 & 0.92 &0.04 \\ 
\hline
3.76524 & 0.0502627 & 0.93 & 0.05 \\ 
\hline
3.73406& 0.0532528 & 0.94 & 0.06 \\
\hline
3.70329 & 0.0564693 & 0.95 & 0.07 \\
\hline
3.65059 & 0.0614452 & 0.96 & 0.085 \\
\hline
3.59917 & 0.0670383 & 0.97 & 0.1 \\
\hline

\end{tabular}
\caption{Estimated values of RKRBH parameters $\ell$ and $a/M$ from known shadow observables $R_s$ and $\delta$ at an inclination angle $\theta=\pi/2$, with $\gamma=0.1$ and $Q=0.5M$.}\label{parameter_table1}
\end{table*}
\begin{table*}

\begin{tabular}{|p{1.8cm}|p{1.8cm}|p{1.8cm}|p{1.8cm}| }
\hline
$A$ & $\mathfrak{D}$ & $a/M$  & $\ell$  \\
\hline\hline 
74.1515&0.0552995 &0.7597  &0.0 \\ 
\hline 
72.3524& 0.0516556 & 0.703 &0.01 \\
\hline
65.2827&  0.0353269 & 0.537 &0.02 \\
\hline 
63.1388 & 0.024084 &0.408  &0.03 \\
\hline 
61.0337& 0.0236297 & 0.399 & 0.04\\
\hline 
58.9738 & 0.0202721 & 0.353 & 0.05\\
\hline 
56.9573& 0.0166066 &0.300  &0.06 \\
\hline 
54.9832 & 0.0135122 & 0.252 &0.07 \\
\hline 
54.4023 & 0.00516638 & 0.1087 &0.073 \\
\hline 
51.1622 & 0.00475311 & 0.0992 &0.09 \\

\hline

\end{tabular}
\caption{Estimated values of RKRBH parameters $\ell$ and $a/M$ from known shadow observables $A$ and $\mathfrak{D}$ at inclination angle $\theta=\pi/2$ with $Q=0.5M$ and $\gamma=0.1$.}\label{parameter_table2}
\end{table*}

\section{The energy emission rate}\label{sec:results}
From a quantum mechanical perspective, particles can be created and annihilated at the event horizon of a black hole. Particles with positive energy may escape this horizon through a process known as tunnelling. This phenomenon gives rise to black hole radiation, which could ultimately lead to the evaporation of the black hole. BH evaporation occurs as a result of Hawking radiation, a quantum phenomenon in which black holes emit thermal radiation, gradually losing mass and energy over time until they ultimately disappear \cite{Hawking:1974rv}. At elevated energy levels, Hawking radiation is typically released within a finite cross-sectional area, referred to as $\sigma_l$. For distant observers located far from the black hole, this cross-sectional area gradually approaches the shadow cast by the black hole \cite{Wei:2013kza}. It has been shown that $\sigma_l$ is directly related to the area of the photon ring, a relationship that can be mathematically expressed as \cite{Wei:2013kza, Decanini:2011xw,Li:2020drn}
\begin{equation}
\sigma_l \approx \pi \bar{R}_{\rm{sh}}^2.
\label{eq:sigmal}
\end{equation}
\begin{figure}[hbt!]
  	\centering
  	\includegraphics[width=9cm,height=7.5cm]{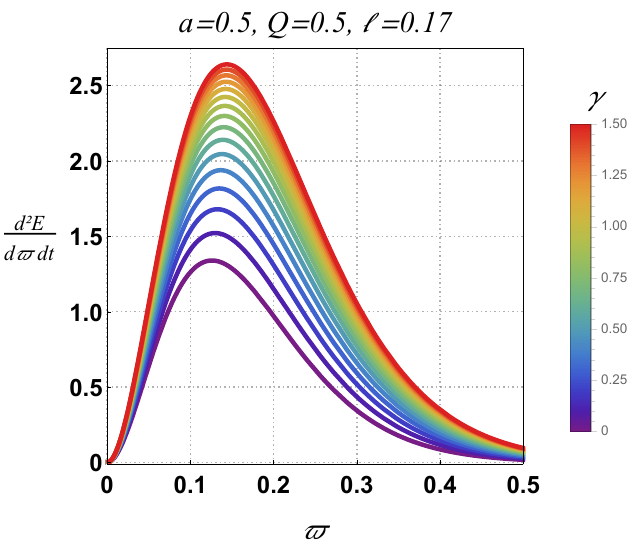}~~~
  	\includegraphics[width=9cm,height=7.5cm]{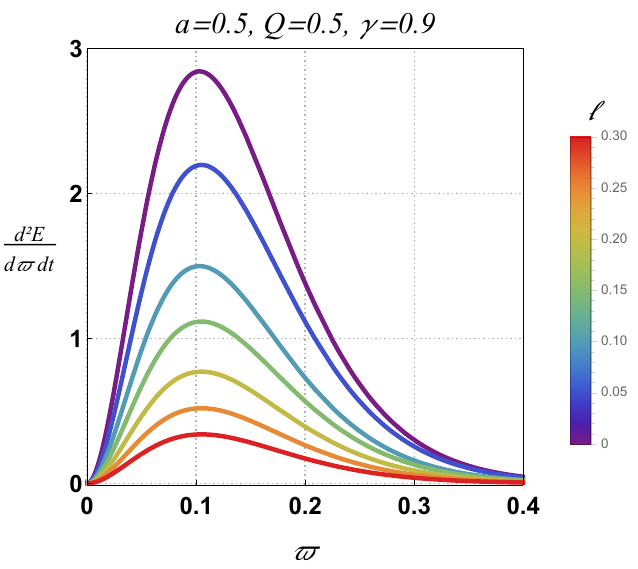}~~~~\\
\includegraphics[width=9cm,height=7.8cm]{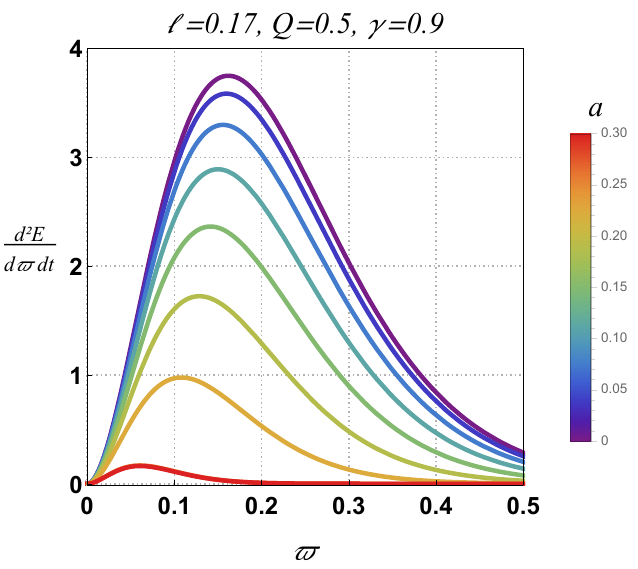}~~~
\includegraphics[width=9cm,height=7.8cm]{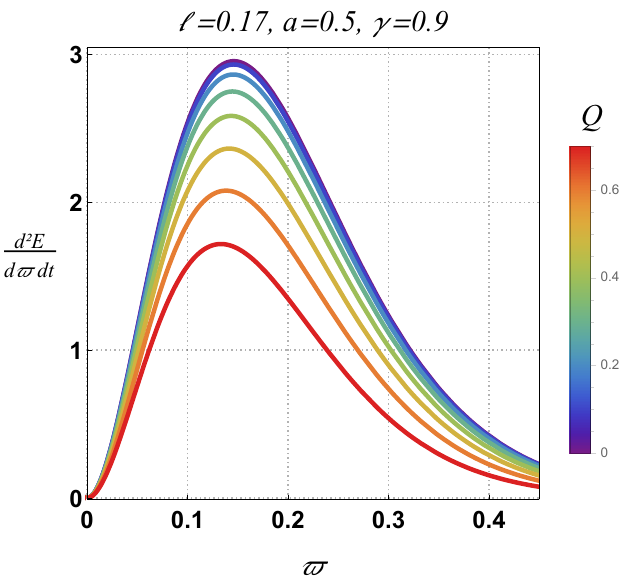}~~~
  	\caption{Evolution of the energy emission rate with the frequency $\bar{\omega}$ for RKRBHs with varying $\gamma$ (top left), $\ell$ (top right), $a$ (bottom left) and $Q$ (bottom right).}
	 \label{Figeer}
\end{figure}
Accordingly, the energy emission rate of the BH can be expressed as
\begin{equation}
\Omega \equiv \frac{d^2 E(\varpi)}{d\varpi \, d t} = \frac{2\pi^2 \sigma_l}{e^{\varpi / T_\mathrm{H}^+} - 1} \varpi^3 \approx \frac{2\pi^3 \bar{R}_{\rm{sh}}^2 \varpi^3}{e^{\varpi / T_\mathrm{H}^+} - 1},
\label{eq:emissionrate}
\end{equation}
where \(\varpi\) is the emission frequency, and \(T_\mathrm{H}^+ = {\kappa}/{2\pi}\) is the Hawking temperature at the event horizon, where
\begin{equation}
\kappa = \left. \frac{\Delta_r'(r)}{2\left(a^2 + r^2\right)} \right|_{r_+},
\label{eq:kappa}
\end{equation}
is the surface gravity at the event horizon. It is straightforward to verify that for zero spin parameter (i.e. \(a = 0, Q=0, \ell=0\)), this quantity reduces to \(\kappa = \mathbb{F}_1'(r_+)/2\), which is the surface gravity of the event horizon for static BHs. In Fig. \ref{Figeer}, some examples of the behaviour of \(\Omega\) versus changes in the frequency \(\varpi\) are plotted for the rotating BH, with BH parameters taken within the observationally accepted domains, as identified in the previous section. Thus, using the geometric-optics proxy $\sigma_{\ell} \simeq \pi \bar{R}_{\text{sh}}^{2}$, the plots show how spectral peaks and high-frequency tails respond to model parameters. 
The Hawking-temperature dependence (through the surface gravity $\kappa$) implies that modifications to the horizon radius induced by $\ell$ and $e^{-\gamma}Q^{2}$ directly shift the emission peak: increasing $\ell$ or $Q$ (with small $\gamma$) tends to increase the surface gravity and shift the peak to higher frequencies, whereas raising $\gamma$ softens charge effects and moves the spectrum closer to the uncharged baseline.

\section{ CONSTRAINTS FROM THE EHT Observation}\label{sec:constraints}
The locus of photon trajectories that escape the gravitational potential delineates the observed black hole shadow and therefore encodes strong-field properties of the spacetime \citep{Jaroszynski:1997bw,Falcke:1999pj}. The EHT images of the supermassive compact objects M87* and Sgr A* \citep{EventHorizonTelescope:2019dse,EventHorizonTelescope:2019ggy,EventHorizonTelescope:2022wkp,EventHorizonTelescope:2022xqj} are broadly consistent with the shadow morphologies expected for Kerr black holes in GR, lending empirical support to the Kerr hypothesis \citep{Psaltis:2007cw}. Nevertheless, owing to residual observational uncertainties, alternative explanations stemming from modified theories of gravity (MTGs) remain viable. Consequently, shadow measurements furnish a valuable probe of possible strong-field quantum corrections and provide direct tests of the Kerr/no-hair paradigm \citep{Carter:1971zc}.

Shadow observables, in particular the shadow area and derived radii-sensitively reflect the detailed structure of the underlying spacetime and therefore offer a route to constrain departures from the Schwarzschild/Kerr geometries. To evaluate the extent to which current EHT resolution can probe Lorentz-violating motivated shadow diameter, we model M87* and Sgr A* as RKRBHs and confront their shadow predictions with the EHT bounds on the Schwarzschild shadow diameter, $\theta_{sh}$.

For a source at distance $d$ the corresponding angular shadow diameter is \citep{Bambi:2019tjh,Kumar:2020owy,Afrin:2021imp}
\begin{equation}\label{angularDiameterEq}
\theta_{sh}=2\frac{R_a}{d}\;,\qquad R_a=\sqrt{A_s/\pi},
\end{equation}
so that $\theta_{sh}$ depends on both $d$ and the central mass $M$ through $A_s$.

Within GR, Kerr shadows may be up to \(\sim 7.5\%\) smaller than the Schwarzschild value; hence the interval \(-0.075\leq\delta\leq0\) is compatible with Kerr geometries as the spin parameter ranges \(0\leq a/M\leq1\) and the observer inclination varies over \(0^\circ\leq\theta_0\leq90^\circ\) \citep{EventHorizonTelescope:2022xqj}. The angular shadow diameter $\theta_{sh}$ is therefore the only important observable that we will consider to constrain the parameters $\ell$ and $Q$, even though the EHT observations offer comprehensive information about the image of M87* \citep{EventHorizonTelescope:2019dse} and Sgr A* \citep{EventHorizonTelescope:2022wkp}.

\subsection{Constraints from the M87* observations}

The EHT observational results from the shadow images of M87* revealed the areal radius and deformation of the shadow, respectively, as $4.31M \leq R_a \leq 6.08M$ and $1 \leq \mathfrak{D} \leq 1.33$~\cite{EventHorizonTelescope:2019dse}. Furthermore, it has been shown that the spin parameter of M87* is approximately $a = (0.9 \pm 0.05)M$~\cite{Tamburini:2019vrf}. Following the discussion in Ref.~\cite{Daly:2023axh}, we assume an observer inclination angle of $\theta_o = 17^\circ$.
Utilizing very-long-baseline interferometry (VLBI) technology, the EHT collaboration obtained the first image of the supermassive BH M87*~\cite{EventHorizonTelescope:2019dse,EventHorizonTelescope:2019ths,EventHorizonTelescope:2019ggy}. The observed shadow image exhibits a nearly circular, crescent-shaped morphology with an emission zone angular diameter of $\theta_{sh} = 42 \pm 3\,\mu\text{as}$, where the $\pm 3\,\mu\text{as}$ represents the observational uncertainty. These results are consistent with simulated images of Kerr BHs \cite{EventHorizonTelescope:2019dse,EventHorizonTelescope:2019ths,EventHorizonTelescope:2019ggy}.

\begin{figure*}[hbt!]
\begin{tabular}{c c}
\includegraphics[scale=0.65]{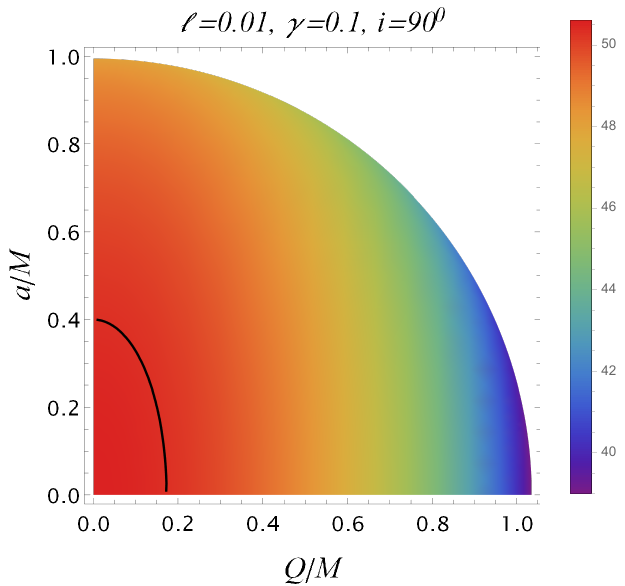}
\includegraphics[scale=0.65]{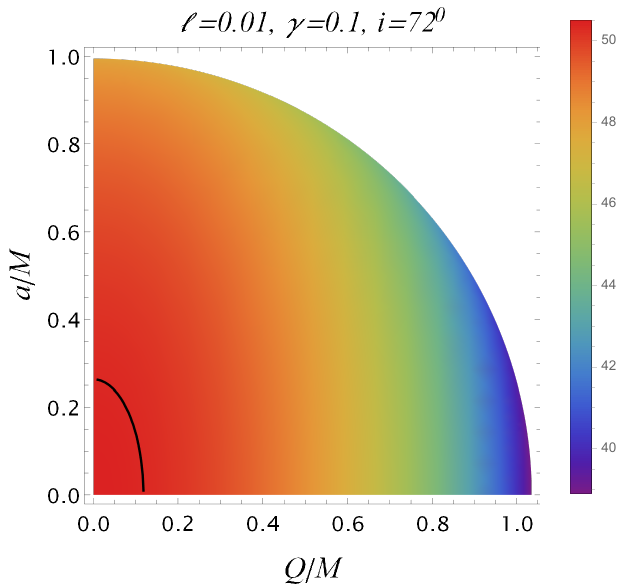}\\
\includegraphics[scale=0.65]{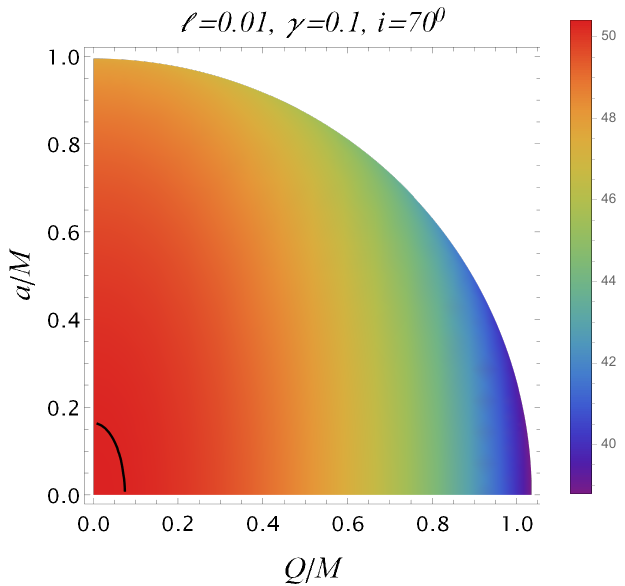}
\includegraphics[scale=0.65]{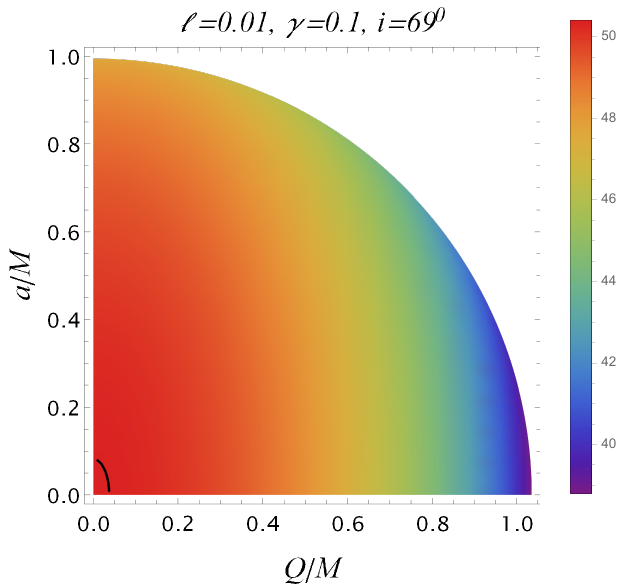}
\end{tabular}	
\caption{Angular diameter observable $\theta_{sh}$, in units of $\mu$as, for RKRBH shadows as a function of parameters ($a/M$, $Q/M$) with black solid curve corresponding to $\theta_{sh}=39\mu$as of the M87* BH. %The substantial region is consistent with the EHT observation of M87*, and the white region is forbidden for ($a/M$, $b_0$). The inclination angle is $\theta_0=17$\textdegree~$(left)$ and $\theta_0=90$\textdegree~$(right)$.
}
\label{FigM87-2}

\end{figure*}

Interpreting EHT observations is challenging due to the sparse telescope array and uncertainties in modeling radiation, plasma, and accretion physics \citep{Gralla:2020pra, Gralla:2019xty}. It is still debated whether the observed rings in M87* and Sgr A* images arise from the accretion flow, the lensed photon region, or both \citep{Gralla:2020pra, Gralla:2019xty}. In this study, we identify the bright emission ring surrounding the dark core as the black hole shadow and compute its observables accordingly. The EHT inferred an inclination of about $163^\circ$ for M87*, based on its jet direction \citep{Walker:2018muw}.

Accounting for a potential emission-related offset of up to  \(10\%\) 
between the observed bright ring and the true photon ring, we adopt the calibrated angular diameter \(\theta_{\mathrm{sh}}=37.8\pm 2.7~\mu\mathrm{as}\) for M87* \citep{EventHorizonTelescope:2019dse,Banerjee:2022bxg}. In the panels of Fig.~\ref{FigM87-2} we display the modelled angular diameter \(\theta_{\mathrm{sh}}\) for our rotating, charged RKRBH family over the \((a/M,Q/M)\) plane evaluated at four representative observer inclinations \(\theta_o\in\{90^\circ,72^\circ,70^\circ,69^\circ\}\). To obtain conservative exclusions we require
\begin{equation}
\theta_{\mathrm{sh}}\;\geq\;37.8~\mu\mathrm{as},
\end{equation}
so that theoretical predictions do not fall below the EHT lower observational bound after allowing for a \(\pm10\%\) emission-offset systematic.

The four inclinations were chosen to illustrate both extremal projection effects and the sensitivity of the inferred parameter-space bands to small changes in inclination. Qualitatively the results are as follows. For the edge-on case (\(\theta_o=90^\circ\)) projection amplifies spin-induced asymmetry while leaving size reductions from electromagnetic/Kalb--Ramond effects clearly visible; consequently the \(\theta_{\mathrm{sh}}\geq37.8~\mu\mathrm{as}\) requirement excludes the largest portion of the \((a,Q)\) plane here  in particular, combinations with large effective charge (see below) coupled to weak ModMax screening are disfavoured. Progressively lowering the inclination to \(\theta_o=72^\circ,70^\circ\) and \(69^\circ\) reduces the apparent asymmetry and increases the apparent diameter for the same intrinsic parameters, so the excluded region shrinks: the most permissive configuration among those shown is \(\theta_o=69^\circ\), for which the same \((a,Q)\) pair that was excluded at \(\theta_o=90^\circ\) can become marginally allowed. This behaviour highlights a practical degeneracy between inclination and intrinsic deviations: a model with larger \(Q\) (or larger \(\ell\)) seen at lower inclination can mimic the \(\theta_{\mathrm{sh}}\) of a less charged model seen edge-on.

Physically, three competing effects control these curves. First, the electric charge \(Q\) reduces the photon capture radius and therefore tends to shrink \(\theta_{\mathrm{sh}}\) for fixed \(M\). Second, the ModMax nonlinearity \(\gamma\) operates as a screening factor \(\propto e^{-\gamma}\) in the electromagnetic contribution, so larger \(\gamma\) mitigates charge-driven shrinkage and relaxes the bound. Third, the Lorentz-violating amplitude \(\ell\) effectively rescales metric coefficients (entering via factors of \((1-\ell)^{-1}\) and \((1-\ell)^{-2}\)) and typically increases the compactness seen by null geodesics, thereby reducing \(\theta_{\mathrm{sh}}\) and tightening the constraint. It is convenient to summarise these combined effects through the single effective combination that governs the leading charge contribution in the metric,
\begin{equation}
\mathcal{Q}_{\rm eff}\equiv\frac{e^{-\gamma}Q^2}{(1-\ell)^2},
\end{equation}
which controls the dominant charge-induced shadow shrinkage. In Fig.~\ref{FigM87-2} the \(\theta_{\mathrm{sh}}=\) const contours therefore shift predominantly along directions of increasing or decreasing \(\mathcal{Q}_{\rm eff}\); \(\gamma\) and \(\ell\) move these contours in qualitatively different (partly orthogonal) manners, producing the degeneracy lines visible in the four panels.

Finally, because the EHT observable is an emission-weighted ring rather than a direct, unambiguous image of the photon ring, robust inference from these panels requires folding in GRMHD forward models and marginalising over ring-offset systematics (we adopt the conservative \(\pm10\%\) offset throughout). The net phenomenological conclusion illustrated by Fig.~\ref{FigM87-2} is that  across the four representative inclinations; a considerable portion of the RKRBH parameter space remains compatible with M87*'s measured diameter, but parameter combinations characterised by large \(\mathcal{Q}_{\rm eff}\) (i.e. large \(Q\), small \(\gamma\), and/or \(\ell\) near unity) are systematically disfavored, with the degree of exclusion growing with observer inclination.

\subsection{Constraints from the Sgr A* observations}
The parameters of Sgr A*, as observed by the EHT, have been further constrained using data from the VLTI and the Keck observatories. These measurements provide bounds on the areal radius and the fractional deviation of the black hole shadow: $4.5M \leq R_a \leq 5.5M$ and $-0.17 \leq \delta \leq 0.01$ from VLTI, and $4.3M \leq R_a \leq 5.3M$ and $-0.14 \leq \delta \leq 0.05$ from Keck \citep{EventHorizonTelescope:2022wkp}. Additionally, based on general relativistic magnetohydrodynamic (GRMHD) simulations with spin values $a = 0.5M$ and $a = 0.94M$, the EHT collaboration inferred an inclination angle $\theta_o < 50^\circ$ consistent with observational constraints. A recent study further refines the spin estimate of Sgr A* to $a = (0.9 \pm 0.06)M$ \citep{Daly:2023axh}.

The Sgr~A* shadow data used here are the EHT 1.3\,mm reconstructions and modelling results \citep{EventHorizonTelescope:2022exc,EventHorizonTelescope:2022urf,EventHorizonTelescope:2022apq,EventHorizonTelescope:2022wok,EventHorizonTelescope:2022wkp,EventHorizonTelescope:2022xqj}. Because Sgr~A* probes far higher curvature scales (for comparable dimensionless parameters) and benefits from an independently determined mass-distance calibration (\(M=4.0^{+1.1}_{-0.6}\times10^6M_\odot\), \(d\simeq 8\) kpc), its angular-diameter measurement provides complementary and in some regimes stronger limits than M87*. The EHT reports a mean angular diameter \(\theta_{\rm sh}=48.7\pm 7~\mu\mathrm{as}\), and inclinations \(\theta_o\gtrsim 50^\circ\) are not preferred by the imaging analyses; nonetheless for consistency with the M87* comparison we evaluate \(\theta_{\mathrm{sh}}\) over the same four inclinations \(\theta_o\in\{90^\circ,72^\circ,70^\circ,69^\circ\}\) and compare with the conservative one-sided requirement
\begin{equation}
\theta_{\mathrm{sh}}\;\geq\;48.7~\mu\mathrm{as}.
\end{equation}

\begin{figure*}[hbt!]
\begin{tabular}{c c}
\includegraphics[scale=0.65]{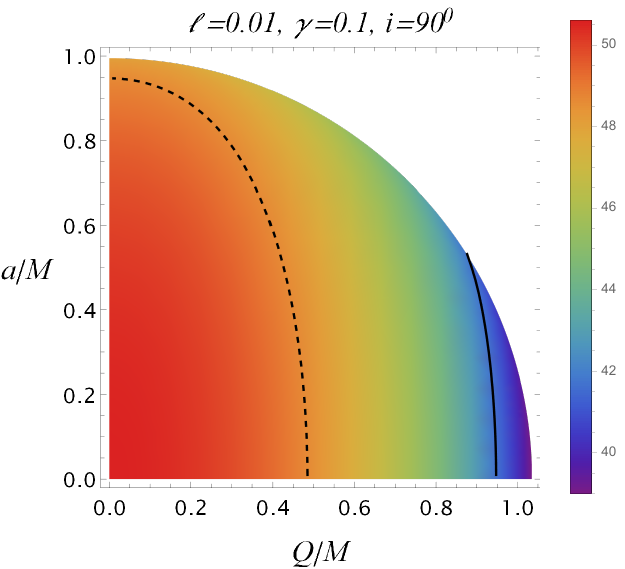}
\includegraphics[scale=0.65]{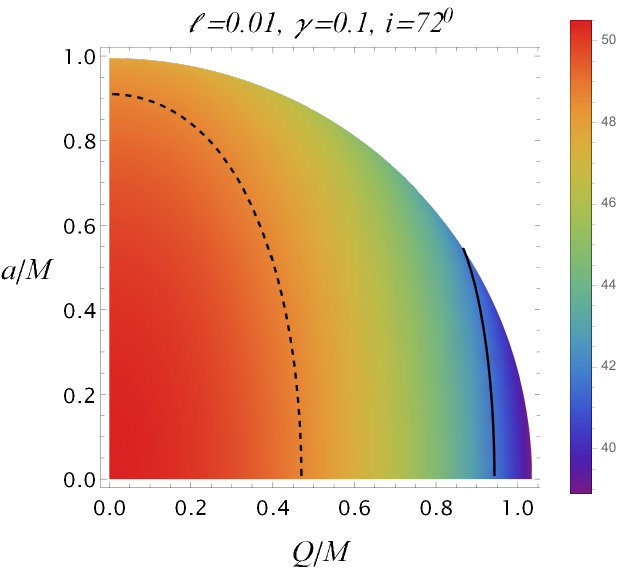}\\
\includegraphics[scale=0.65]{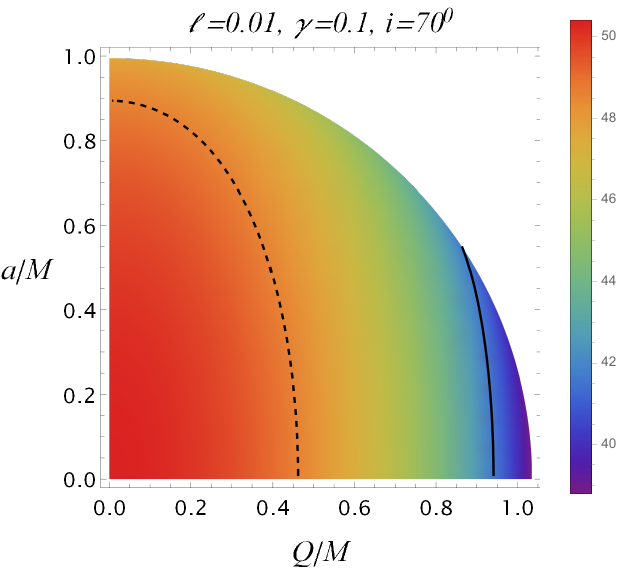}
\includegraphics[scale=0.65]{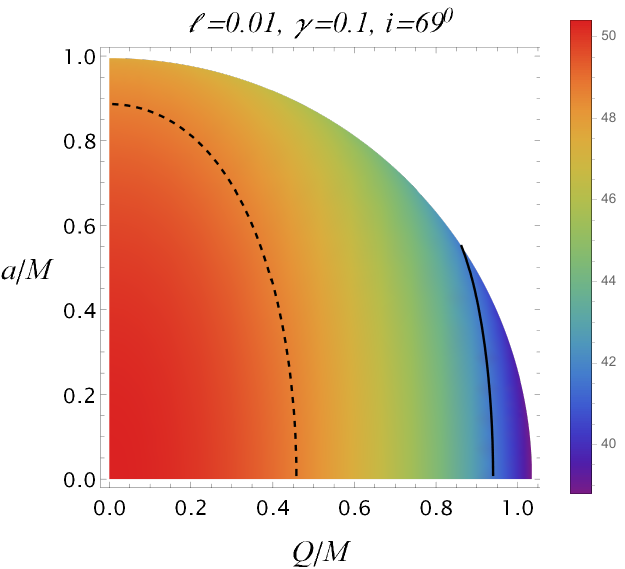}
\end{tabular}	
\caption{Angular diameter observable $\theta_{sh}$, in units of $\mu$as, for RKRBH shadows as a function of parameters ($a/M$, $Q/M$) with black solid curve corresponding to $\theta_{sh}=41.7\mu$as and black dashed curve corresponding to $\theta_{sh}=48.7\mu$as  of the SgrA* BHs. 
%The substantial region is consistent with the EHT observation of SgrA*, and the white region is forbidden for ($a/M$, $b_0$). The inclination angle is $\theta_0=50$\textdegree~$(left)$ and $\theta_0=90$\textdegree~$(right)$.
}\label{FigSgrA}
\end{figure*}

The Sgr~A* panels in Fig.~\ref{FigSgrA} show qualitatively similar trends to M87* but with two important differences. First, because the angular-diameter uncertainty is larger in absolute terms relative to the baseline ($\pm 7\,\mu$as), the fraction of parameter space formally compatible with the conservative lower bound is larger; second, the independent mass-distance calibration reduces a class of degeneracies that affect M87*, so constraints translated into \(\mathcal{Q}_{\rm eff}=e^{-\gamma}Q^2/(1-\ell)^2\) are somewhat more robust when expressed as dimensionless bounds. For the four inclinations explored here the exclusion power is again strongest at \(\theta_o=90^\circ\) and weakens as the inclination decreases to \(72^\circ,70^\circ\) and \(69^\circ\), with the consequence that parameter combinations mildly disfavored at \(\theta_o=90^\circ\) can become permissible at \(\theta_o=69^\circ\). In particular, combinations with very large \(\mathcal{Q}_{\rm eff}\) are excluded across most inclinations, while modest effective charges (achieved either via small \(Q\), significant ModMax screening \(\gamma\), or moderate \(\ell\)) remain allowed.

We also compare our results with the EHT-reported Schwarzschild-deviation intervals for Sgr~A* (Keck and VLTI estimates). Mapping those deviation bands into our RKRBH parameter space via metric-dependent ray-tracing shows that the allowed slices are compatible with the conservative angular-diameter bound quoted above: i.e. the region of \((a,\mathcal{Q}_{\rm eff})\) consistent with the Keck/VLTI \(\delta_{SgrA^*}\) intervals lies largely inside the \(\theta_{\mathrm{sh}}\geq48.7~\mu\mathrm{as}\) region once the full traceless-field and ModMax effects are accounted for. We emphasise that the traceless (conformal) matter sector modifies the angular dependence of the lensing potential (and hence the mapping between \(\delta\) and \(\mathcal{Q}_{\rm eff}\)), so that translating EHT-derived \(\delta\)-bounds into constraints on \((a,\gamma,\ell,Q)\) requires the same ray-tracing pipeline used to generate the figures shown here.

In summary, the joint M87*/Sgr~A* comparison across \(\theta_o\in\{90^\circ,72^\circ,70^\circ,69^\circ\}\) demonstrates: (i) inclination is a key degeneracy direction and small changes in \(\theta_o\) noticeably shift allowed bands, (ii) ModMax screening \(\gamma\) and the Lorentz-violating parameter \(\ell\) affect the shadow size in qualitatively distinct ways (hence multi-observable inference can break degeneracies), and (iii) under conservative assumptions RKRBH geometries with modest \(\mathcal{Q}_{\rm eff}\) remain viable descriptions of both EHT targets, whereas large \(\mathcal{Q}_{\rm eff}\) (large \(Q\), small \(\gamma\), \(\ell\!\to\!1\)) are disfavoured - the degree of disfavour depends monotonically on the adopted inclination from \(\theta_o=69^\circ\) (weakest) to \(90^\circ\) (strongest). 

Comparison of the two figures (\ref{FigM87-2}-\ref{FigSgrA}) makes clear that Sgr A's larger fractional diameter uncertainty and intrinsic variability produce broader allowed bands (less restrictive size-only bounds) even though its independently calibrated mass-distance reduces degeneracies when results are expressed in dimensionless form. Thus Sgr~A* is currently more limited by imaging systematics/variability, whereas M87's smaller fractional uncertainty yields tighter imaging-only exclusions.

\section{Discussion and conclusion}\label{sec:discussion}
In this work we have presented a comprehensive, observation oriented analysis of rotating, charged black holes in a ModMax-Kalb-Ramond traceless-conformal matter framework (RKRBH). By combining analytic inspection of the metric and horizon structure with high-resolution ray-tracing and confrontation with EHT measurements for M87* and Sgr~A*, we have established a suite of theoretical and phenomenological results that clarify how the ModMax parameter \(\gamma\), the Kalb-Ramond amplitude \(\ell\), the electric charge \(Q\), and the traceless (conformal) matter sector jointly shape strong-field observables.

Analytically, the leading electromagnetic contribution to the metric appears through the effective combination
\[
\mathcal{Q}_{\mathrm{eff}}\equiv\frac{e^{-\gamma}Q^2}{(1-\ell)^2},
\]
so that ModMax nonlinearity produces an \(\mathrm{e}^{-\gamma}\) screening of the electric charge while the Lorentz-violating parameter \(\ell\) rescales metric coefficients by powers of \((1-\ell)^{-1}\). As a direct consequence, increasing \(Q\) (for fixed \(\gamma,\ell\)) tends to reduce the outer horizon radius \(r_{+}\) and drive the solution toward extremality; increasing \(\ell\) amplifies this effective-charge influence through \((1-\ell)^{-2}\), producing further horizon shrinkage; increasing \(\gamma\) counteracts charge effects by reducing \(\mathcal{Q}_{\mathrm{eff}}\) and tends to restore larger horizon radii. These competing dependencies reshape the extremality separatrix and determine the allowed regions of parameter space that admit regular black hole solutions.

Ray-traced photon capture contours and shadow morphology are sensitive to the three-way interplay \((\gamma,\ell,Q)\) and to observer inclination \(\theta_{o}\). Electric charge primarily reduces the photon capture radius and hence the shadow angular diameter \(\theta_{\mathrm{sh}}\). ModMax screening (\(\gamma\)) mitigates this shrinkage, producing larger and more circular shadows for fixed intrinsic charge. The KR parameter \(\ell\) imprints qualitatively distinct angular-dependent distortions because it modifies metric prefactors and the angular structure of the lensing potential; thus \(\ell\) affects not only shadow size but also higher-order shape diagnostics (centroid shift, distortion \(\delta\), oblateness \(D\)). The traceless (conformal) matter sector further adjusts the angular dependence of the stress-energy and the near-horizon effective potentials, so that the mapping between metric parameters and observable shadow moments is nontrivial and metric-dependent ray-tracing is required for faithful translation of observational bounds.

The theoretical implications of this work are twofold. First, the RKRBH family provides an explicit phenomenological bridge between non-linear electrodynamics (ModMax), Kalb-Ramond-induced metric rescalings, and observationally accessible strong-field probes; the effective parameter \(\mathcal{Q}_{\mathrm{eff}}\) makes this connection manifest. Second, the traceless (conformal) matter sector shows that conformal-like couplings can leave measurable imprints on strong-lensing observables, motivating further study of traceless stress-energy contributions in alternative-gravity phenomenology.

In summary, the ModMax-Kalb-Ramond traceless-field extension of rotating charged black holes produces clear, model specific imprints on horizon structure, photon capture radii, shadow morphology and-qualitatively-on emission spectra. When confronted with current EHT measurements of M87* and Sgr~A*, RKRBH geometries with modest effective charge \(\mathcal{Q}_{\mathrm{eff}}\) (achieved through small \(Q\), significant ModMax screening \(\gamma\), or moderate \(\ell\)) remain fully compatible with the data, while large \(\mathcal{Q}_{\mathrm{eff}}\) (large \(Q\), small \(\gamma\), \(\ell\!\to\!1\)) are progressively disfavoured; the strength of exclusion increases with inclination from \(\theta_o=69^\circ\) (weakest) to \(\theta_o=90^\circ\) (strongest). These results establish a concrete observationally testable mapping between non-linear electrodynamics, traceless-field couplings, and black-hole imaging, and they set the stage for refined constraints with next-generation VLBI and multi-messenger data.

%-------------------------------------------------------------
%\begin{acknowledgments}

%\end{acknowledgments}

\bibliography{reference.bib}
\bibliographystyle{apsrev4-1}

\end{document}